\documentclass[twocolumn,times]{aastex63}
\usepackage{amsmath}
\usepackage{amssymb}
\usepackage{multirow}
\usepackage{afterpage}
\usepackage{xspace}

\newcommand\gaia{\emph{Gaia}\xspace}
\newcommand\allwise{\emph{AllWISE}}
\newcommand\MSun{\ensuremath{\rm{M}_\sun}}
\newcommand\kms{\ensuremath{\rm{km\,s^{-1}}}}
\newcommand\gcs{\ensuremath{\rm{g\,cm^{-2}}}}
\newcommand\aeaur{AE~Aur}
\newcommand\mucol{$\mu$~Col}
\newcommand\iotaori{$\iota$~Ori}

\newcommand\mualphastar{\ensuremath{\mu_{\alpha*}}}

\newcommand\masyr{ \ensuremath{\rm mas\,yr^{-1}}}
\renewcommand\deg{\ensuremath{\degr}}
\newcommand\dmin{\ensuremath{D_{\rm min}}}
\newcommand\tback{\ensuremath{ t_{\rm back}}}
\newcommand\Ntraceback{16,994\xspace}
\newcommand\Nsys{\ensuremath{N_{\rm s}}}
\newcommand\gband{\ensuremath{G}}
\newcommand\gflux{\ensuremath{F_G}}
\newcommand\positionAngle{\ensuremath{\theta_{\rm PA}}}

\shorttitle{Hunting for Runaways from the Orion Nebula Cluster}
\shortauthors{Farias, Tan \& Eyer}

\begin{document}

\title{\Large Hunting for Runaways from the Orion Nebula Cluster}

\correspondingauthor{Juan P. Farias}
\email{juan.farias@chalmers.se}

\author{Juan P. Farias }
\affil{Dept.\ of Space, Earth \& Environment, Chalmers University of
Technology, Gothenburg, Sweden}

\author{Jonathan C. Tan}
\affil{Dept.\ of Space, Earth \& Environment, Chalmers University of
Technology, Gothenburg, Sweden}
\affil{Dept.\ of Astronomy, University of Virginia, Charlottesville,
VA 22904, USA}

\author{Laurent Eyer}
\affil{Geneva Observatory, University of Geneva,  Chemin des Maillettes 51, CH-1290,
Versoix, Switzerland}

\begin{abstract}
We use \gaia\ DR2 to hunt for runaway stars from the Orion Nebula Cluster (ONC). We
search a region extending 45$\deg$\ around the ONC and out to 1~kpc to find sources
that overlapped in angular position with the cluster in the last $\sim$10~Myr. We
find $\sim17,000$ runaway/walkaway candidates satisfy this 2D traceback condition.
Most of these are expected to be contaminants, e.g., caused by Galactic streaming
motions of stars at different distances. We thus examine six further tests to help
identify real runaways, namely: (1) possessing young stellar object (YSO) colors
and magnitudes based on \gaia\ optical photometry; (2) having IR excess consistent
with YSOs based on 2MASS and WISE photometry; (3) having a high degree of optical
variability; (4) having closest approach distances well constrained to within the
cluster half-mass radius; (5) having ejection directions that avoid the main
Galactic streaming contamination zone; and (6) having a required radial velocity
(RV) for 3D overlap of reasonable magnitude (or, for the 7\% of candidates with
measured RVs, satisfying 3D traceback).  Thirteen sources, not previously noted as
Orion members, pass all these tests, while another twelve are similarly promising,
except they are in the main Galactic streaming contamination zone. Among these 25
ejection candidates, ten with measured RVs pass the most restrictive 3D traceback
condition.  We present full lists of runaway/walkaway candidates, estimate the
high-velocity population ejected from the ONC and discuss its implications for
cluster formation theories via comparison with numerical simulations. 
\end{abstract}


\section{Introduction}\label{sec:intro}

The formation of star clusters takes place inside and from dense molecular clouds
and is the result of gravitational collapse and fragmentation. In these early
stages, young stellar clusters can be dense and the number of stars per cubic
parsec increases as more stars are formed. Strong interactions between stars are
expected to take place relatively frequently, especially if a significant fraction
of stars are formed as binaries or higher-order multiples
\citep[e.g.,][]{Blaauw1961,Poveda1967,Perets2012}.
Such interactions can then lead to the ejection of stars from the cluster at a
range of velocities \citep[e.g.,][]{Fujii2011,Oh2016}. 

The Orion Nebula Cluster (ONC) is the closest massive, dense stellar cluster that
is still undergoing formation, which makes it a perfect laboratory for testing star
cluster formation theories. The ONC's distance is $403\pm7$\,pc \citep{Kuhn2019}.
The total stellar mass is estimated to be $\sim$3000\,$M_\odot$ within a 3\,pc
radius, mixed together with an approximately similar mass of gas in this volume
\citep{DaRio2014}. However, the stars are more centrally concentrated, with their
density dominating inside about 1.4\,pc, having a density profile of $\rho_*\propto
r^{-2.2}$ extending down to $\lesssim$0.1\,pc, where the stellar density reaches
$\gtrsim 10^4\,M_\odot\,{\rm pc}^{-3}$ \citep{DaRio2014}.

Relatively large age spreads have been claimed to be present in the ONC
\citep{Palla2005,Palla2007,DaRio2010,DaRio2016}, suggesting that it has been
forming for the last $\sim$4\,Myr or perhaps even longer. Three different,
relatively discrete sequences in the colour-magnitude diagram have also been
identified \citep{Beccari2017,Jerabkova2019}, which have been interpreted as bursts of star
formation interrupted by the formation and subsequent dynamical self ejection of a
few massive stars \citep{Kroupa2018,Wang2019}. It is expected, however, that massive stars
will also eject a number of lower-mass stars. It has also been shown that the
frequency and velocity distribution of ejected stars is linked to the densest state
in the history of the star cluster \citep{Oh2016}, as well to the dynamical
timescale of its formation \citep{Farias2019}.  Therefore, characterising the
unbound population of a forming star cluster can provide important constraints for
star cluster formation theories, including helping characterise the formation
history of the system \citep[see, e.g.,][]{Tan2006,Tan2006a}.

The core of the ONC, also known as the Trapezium, is the densest and most
dynamically active region of the cluster. Several runaway stars have been suggested
to have been launched from this region. Using proper motions from the
\emph{Hipparcos} mission, the seminal work of \citep{Hoogerwerf2001} found it
highly likely that the runaway O stars \mucol\ and \aeaur\ were ejected about
2.5\,Myr ago from the ONC, confirming the hypothesis first made by
\cite{Blaauw1954}. Both of these stars remain as the oldest and furthest candidate
runaways to have been ejected from the ONC. High proper motion sources have also
been identified in the more immediate surroundings of the Trapezium. For example,
the Becklin-Neugebauer (BN) object \citep{Becklin1967} is moving at $\sim30\,\kms$
and has been proposed to be ejected either from the $\theta^1$ Ori C binary system
in the Trapezium \citep{Tan2004,Chatterjee2012}, or as part of a multiple system
decay involving radio sources I and a third member \citep{Bally2005,Rodriguez2005}
that is still in the Trapezium cluster region.  Recent observations suggest that
this other source is the nearby radio source x \citep{Luhman2018}, which is moving
at $\gtrsim50\,\kms$ \citep[see also][]{Farias2018,Bally2020}. 

More recently, using the high astrometric precision of \gaia, \cite{McBride2019}
searched an ONC membership list and identified 9 high proper motion stars. All
these would have been ejected relatively recently, within the last 0.4\,Myr ago.
However, given that the ONC has been actively forming stars for perhaps $\sim10$
times longer, it is expected that many other high velocity runaway stars have been
ejected and escaped to greater distances.  However, finding such runaways becomes
challenging as they mix with high proper motion field stars. Logically, most
photometric efforts to identify ONC members have been focused close to the ONC.
Therefore, fast runaway stars ejected more than 0.5\,Myr ago and with velocities
higher than 20\,\kms\ that are at least 10\,pc away (i.e., $\gtrsim$1.5$\deg$) have
likely been missed by such studies. Therefore, most of the identified ONC runaway
stars are still close to the ONC region, with the notable exception of \aeaur\ and
\mucol. 

Just before submission of our paper, we have become aware of a preprint by
\cite{Schoettler2020}, who explored a wider area of 100\,pc around the ONC, tracing
back candidates using projected 2D trajectories classifying candidates by their
ages which were estimated using the \emph{PARSEC} isochrones from
\cite{Bressan2012}. They have found 31 runaways and 54 walkaway candidates.
However, they were very strict on the constraints used, i.e., by assuming an upper
limit of 4\,Myr for the ONC and discarding sources with 2D traceback times longer
than that. They also discarded sources for which \emph{PARSEC} isochronal ages were
shorter than their traceback times.

In this work, we attempt to identify potential ONC runaway candidates using the
unprecedented accuracy and scope of the astrometric measurements of \gaia\
\citep{GaiaSolution2018,GaiaCollaboration2016}.  We explore a large area of 45
degrees around the ONC, which contains $>120$ million sources in the \gaia\
catalog.  Selecting a subset that have more accurately determined proper motions
and that are within 1\,kpc distance of the Sun, we trace back the two dimensional
trajectories of such stars. We also test candidates under different criteria, i.e.,
``flags'', in order to identify signatures of youth and astrometric reliability. We
combine this flag system to obtain a smaller number of most probably runaway
candidates. For the small fraction of the sources with radial velocity
measurements, we also consider the more stringent three dimensional trace-back
condition. In this way, we have produced a list of interesting sources that are
prime targets for follow-up observations, e.g., to confirm if they have stellar
properties consistent with ONC membership. We also discuss how the number of prime
candidates that we have found compares to theoretical expectations from cluster
formation simulations. 

\section{ONC frame of reference}\label{sec:oncref}

\begin{figure} 
        \centering
        \includegraphics[width=\columnwidth]{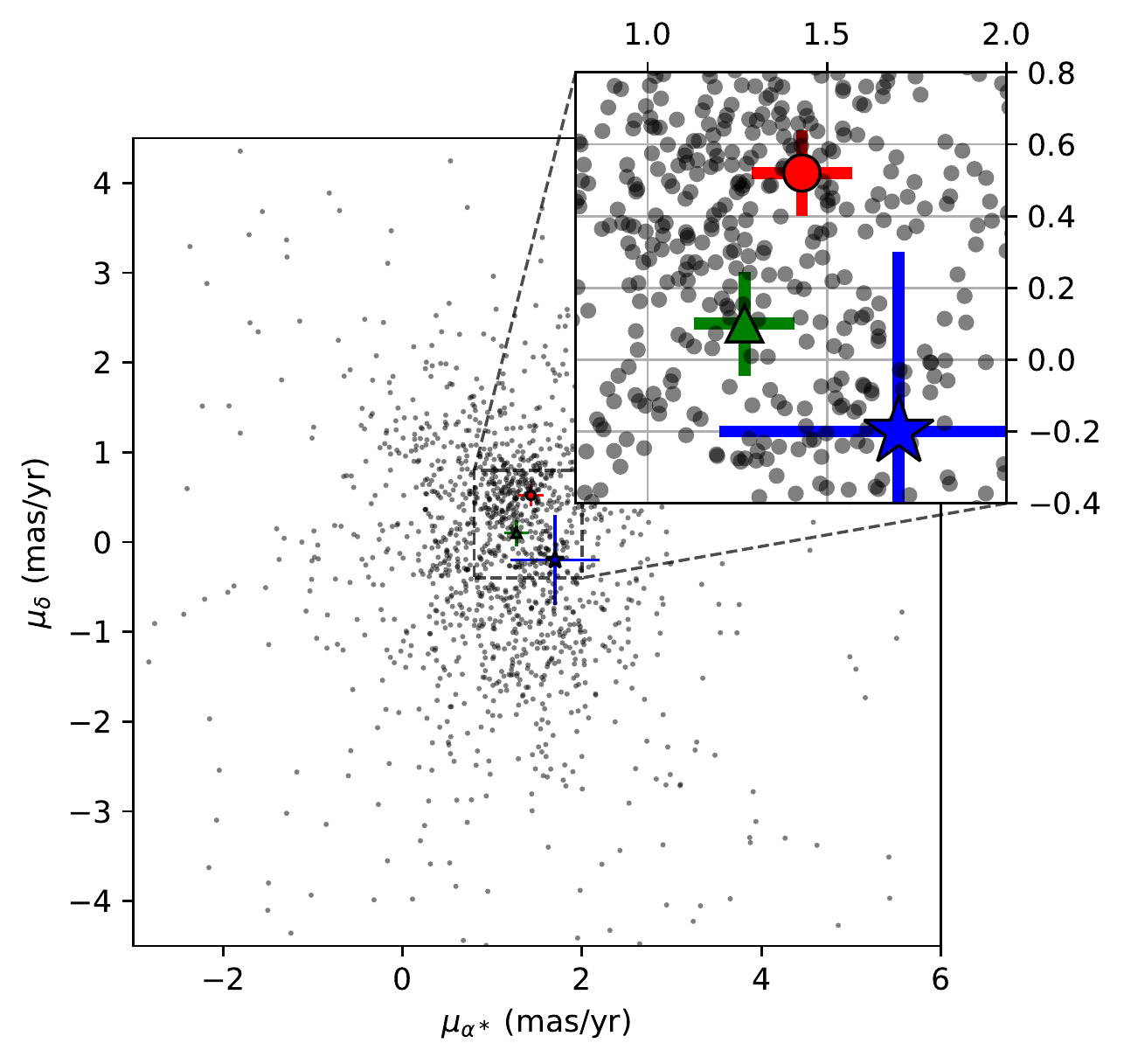}
        \caption{Proper motions of stars thought to be ONC members.  The red circle
                shows our adopted proper motion for the ONC, which is obtained by
                taking a weighted median of the motions of ONC members.  The blue
                star shows the proper motion of the center of mass of \aeaur,
                \mucol\ and \iotaori\ that are proposed to have been ejected from
                the ONC \citep{Hoogerwerf2001}, where the errorbar is an
                approximate estimate based on their distributions.
                The green triangle shows the center of mass proper motion using the
                membership candidates of
                \citet{DaRio2017}.}
        \label{fig:oncref} 
\end{figure}

The starting point for our study is to estimate the proper motion of the ONC. A
good hint of its proper motion is given by the already known runaway stars that
most likely came from the ONC. Two well studied runaway stars are \aeaur\ and
\mucol. They were first noted by \citet{Blaauw1954}, who observed that they travel
with similar velocities ($\sim100\,\kms$) in almost opposite directions. Later,
\citet{Gies1986a} suggested that both stars were ejected from an event that
involved the binary star \iotaori\ \citep[see
also][]{Bagnuolo2001,Gualandris2004,Ryu2017a}. \citet{Hoogerwerf2001} performed a
set of numerical simulations, tracing back the trajectories of \aeaur\ and \mucol\
taking into account the gravitational potential of the Galaxy.  They found that the
proper motion and coordinates of the star cluster that may have hosted the event
are consistent with those of the ONC.

We now estimate the proper motion of the ONC from \gaia\ data and check how
consistent it is with the results reported by \citet{Hoogerwerf2001}. To do so, we
make use of the membership compilation performed by \citet{DaRio2016}.
We cross matched this membership list with the \gaia\ catalog using the best
neighbor method with a 1\arcsec\ cross match threshold, selecting stars within
9\arcmin\ from the ONC center. This angular distance corresponds to the half mass
radius (1\,pc) of the ONC \citep{DaRio2014}.

One of the main problems when estimating the proper motion of the ONC is the
sampling. It would be tempting to use the accurate parallax measurements of \gaia\
to constrain the sample in the line of sight direction on an equivalent distance to
the 9\arcmin\ used as angular distance threshold. However, if we do so the number
of selected stars from the membership list is only on the order of dozens. Such
proper motion would be less reliable since it would be affected by incompleteness
and sparse sampling, since the number of members of the ONC is $\sim3000$
\citep{DaRio2014,DaRio2016}. Therefore, we have opted to use all stars flagged as
members within 9\arcmin\ of the ONC center, which resulted in 458 member stars.

The membership list compiled by \citet{DaRio2016} also includes mass estimates of
the sources, and again, it is tempting to calculate the center of mass proper
motion of the selected sample. However, this quantity can be sensitive to anomalous
motions of small numbers of massive stars, e.g., the most massive star system
$\theta^1C$, which has a relatively high motion within the ONC
\citep[e.g.,][]{Tan2004}. Thus we have also measured the ONC's proper motion using
a weighted median estimate. The weighted median is a robust central tendency
estimator that allows the use of the individual uncertainties in proper motion.
Weights are taken as 1/error$^2$, and then normalized by the total sum of the
weights. As in the median measurement, values are sorted by proper motion and
weights are normalized by its sum. The proper motion at which the cumulative sum of
the normalized weights is 0.5 is then the weighted median.

Figure~\ref{fig:oncref} shows the weighted median proper motion of the ONC members
(red circle, hereafter adopted as the best estimate of ONC proper motion). Errors
were estimated using bootstrap analysis, following the method of \cite{Kuhn2019}.
The resulting measurements are: $\mualphastar = 1.43\pm0.14$\,\masyr\ and
$\mu_{\delta}=0.52\pm0.12$\masyr.  In the figure we also show, for reference, the
proper motion of the center of mass of \aeaur, \mucol\ and \iotaori\ estimated by
\citet{Hoogerwerf2001} (blue star). The three proper motions are in
approximate agreement. Our measurement agrees well with the estimate
of \citep{Kuhn2019}, who measured $\mualphastar = 1.51\pm0.11$\,\masyr\ and
$\mu_{\delta}=0.50\pm0.12$\masyr\ using the same method, but based on a sample from the MYStIX
\citep{Broos2013} and SFiNCs \citep{Getman2017} surveys.

Following the same methodology, we have measured the weighted median radial
velocity of the ONC. For this purpose we have used radial velocity measurements
from the INfrared Spectra of Young Nebulous Clusters (IN-SYNC) survey that covered
the Orion A complex \citep{DaRio2017} obtaining radial velocities for 2691 sources
with uncertainties in individual measurements often being $\lesssim1\,\kms$. Using
this catalog instead of \gaia\, we greatly increase the sample of sources with
available radial velocities within 9\arcmin\ of the ONC, from 15 sources in \gaia\
to 200 sources marked as members of the ONC by \cite{DaRio2016}.  Computing the
weighted median on this sample we have obtained a radial velocity of
$26.4\pm1.6\,\kms$. This radial velocity is consistent with the one that
\cite{Hoogerwerf2001} have estimated for the parent star cluster of \aeaur\ and
\mucol\ of 27.6--28.3\,\kms. In the local standard rest, this radial velocity
transforms into 9.2\,\kms. Such a radial velocity, although somewhat higher than
overall average of stars in the Orion A complex of 8\,\kms\ in \cite{DaRio2017}, is
very consistent with that estimated from $^{13}$CO(2-1) measurements
\citep{Nishimura2015} at the declination of the ONC \cite[see Figure 4, 4th panel
in][]{DaRio2017}, where a radial velocity gradient is shown increasing from low to
higher declinations.

Using the above estimate for the proper motion of the ONC and tracing back the
trajectories of \aeaur\ and \mucol\ using great circles trajectories and assuming
constant proper motions, we find that their closest approaches to the ONC are
22$\pm$28\arcmin\ and 44$\pm$42\arcmin\ respectively.  These two sources are the
furthest known runaway stars from the ONC. The error estimation of their closest
approach comes from the errors in the proper motions. Systematic errors, such as
the neglect here of the Galactic potential, will also contribute. The effects of
both of these types of error grow with the traceback distance. Therefore, for
sources at similar (angular) distances of \mucol\ and \aeaur\ from the ONC, we
expect that true runaways will also exhibit similar errors in their closest
distance to the ONC. Below, we will design our trace-back thresholds in order to
capture \mucol\ and \aeaur.

\section{Sample selection by 2D Traceback}\label{sec:sample}

We first select sources that are up to 45$\deg$\ from the center of the ONC. Within
this region there are 122,531,450 sources in the \gaia\ DR2 catalog. We note that this
surveyed region encloses a sphere of radius $285$\,pc around the ONC, which
contains \aeaur, \mucol\ (and most of the higher likelihood runaway candidates identified;
see below).   
We then used the following constraints to clean the sample. 

First, to limit ourselves to stars with well-behaved astrometric solutions, we
select those sources with reference unit weight error (RUWE) parameter $< 1.4$.
This leaves 108,990,887 sources in the sample. Next, we select stars with parallax
($\varpi$) errors that are $<20\%$, i.e., $\varpi/\sigma_{\varpi} > 5$. This
reduces the sample to 18,118,187 sources.  To carry out a standard variability
study (described later), we require sources to have
\texttt{visibility\_periods\_used} $> 6$. This makes only a minor difference to the
sample size, leaving 18,113,350 sources. Finally, we restrict to sources up to 1
kpc distance (given the ONC's distance of $403$\,pc), which leaves a final sample
of 6,760,924 sources. We note that we have deliberately avoided to impose a
photometric condition to clean the sample using 
\texttt{phot\_bp\_rp\_excess\_factor}, given that there are many reasons why this 
factor may be high and would not directly affect the quality of astrometric 
solution. In any case, only a small fraction of the sample are affected by this 
parameter.

\begin{figure}
    \centering
    \includegraphics[width=\columnwidth]{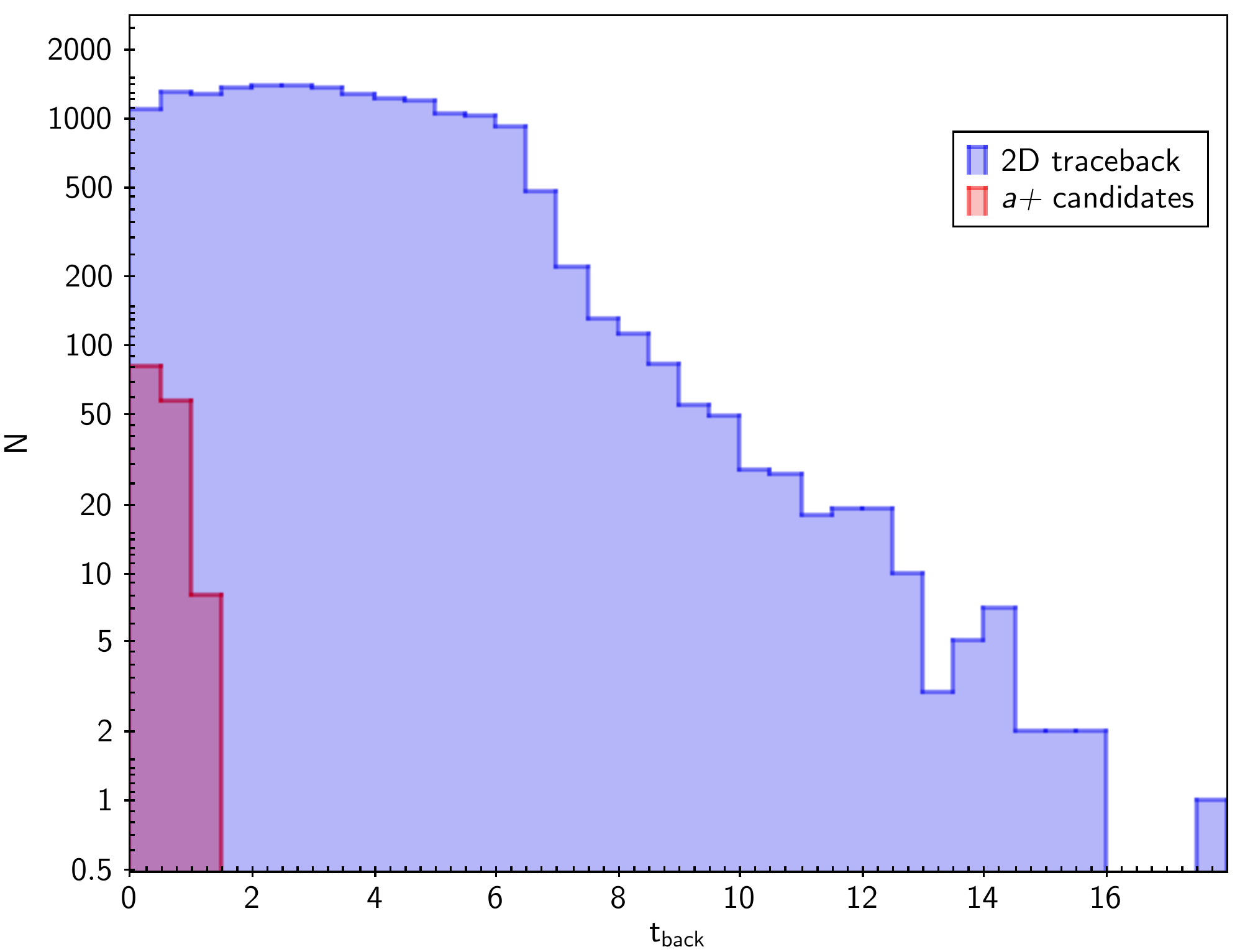}
    \caption{Histogram of obtained traceback times for the $\sim$17,000 sources
    selected with the linear 2D traceback method (Blue) and for the best scored
    candidates (red) with bins of 0.5\,Myr. }%
    \label{fig:tback}
\end{figure}

After defining our sample, we wish to select new runaway candidates using variables
that are common for all stars in the \gaia\ DR2 database. This means that we must
ignore radial velocities for now, since, in our final sample, only $\sim7\%$ of
stars have radial velocity measurements. We will do a further selection with this
small subset at the end of the main analysis.

With the final sample of almost 7 million sources, we then select stars whose
trajectories overlap with that of the ONC in space and time. We use the following
procedure. We remove the Sun's peculiar motion
relative to the Local Standard Rest (LSR), using values from
\citep{Schonrich2010}. We calculate the trajectory of each star and the ONC using
assuming constant proper motion along great circle trajectories on the sky, i.e.,
ignoring effects of acceleration due to the Galactic potential.  Each point on the
trajectory has an associated trace-back time ($\tback$) and we use this to
calculate the closest approach to the ONC ($\dmin$) in space and time. We require
that $\dmin$ is smaller than a certain threshold condition given by
\begin{eqnarray}
\frac{\dmin}{1'}  <  10 + 1.3 \frac{\theta}{1\deg},
\end{eqnarray}
where $\theta$ is the current angular distance of the star from the ONC. Thus the
threshold becomes larger for stars that are currently further away (in angular
distance) from the ONC, which allows for the fact that the errors in estimating
past positions grow with longer traceback distances, i.e., due to proper motion
errors, the constant proper motion approximation (which is broken by projection
effects) and the effects of the Galactic potential. The normalization of the
threshold condition has been adjusted to make sure that \mucol\ and \aeaur\ are
recovered by this method: in particular to capture \mucol\ which has a closest
approach $\dmin = 43\arcmin$ with $\theta=27\deg$. For sources currently close to
the ONC, the threshold is $\sim10\arcmin$, which is about 1.2\,pc, i.e., similar to
the half-mass radius of the cluster. This choice is motivated since runaways are
expected to be produced by dynamical ejection events that are more frequent in the
dense, inner regions of the cluster, and having a smaller threshold helps to
minimize contamination from field stars. 
We note that our method will select all stars that are currently within 10\arcmin\ 
of the ONC's center.

Using this trace-back method we find there are \Ntraceback\ sources that meet this
2D projected overlap condition. We considered adopting a maximum traceback time,
e.g., $\sim5$ to 10\,Myr, however, when we examine the distribution of traceback
times of the selected sources (Figure~\ref{fig:tback}), we see that most sources
are already within this range. Given the caveats of the assumed linear trajectory,
constant proper motion approximation and thus possible discrepancies between 2D and
3D traceback, the real traceback time may be very different especially for sources
that are far from the ONC. Thus, we simply retain all the selected sources for
further analysis and note that the value of the traceback time, especially if
$\gtrsim 5$\,Myr, could weigh against the likelihood of a source being a genuine
runaway from the ONC.

\begin{figure*}
    \centering 
    \includegraphics[width=0.95\textwidth]{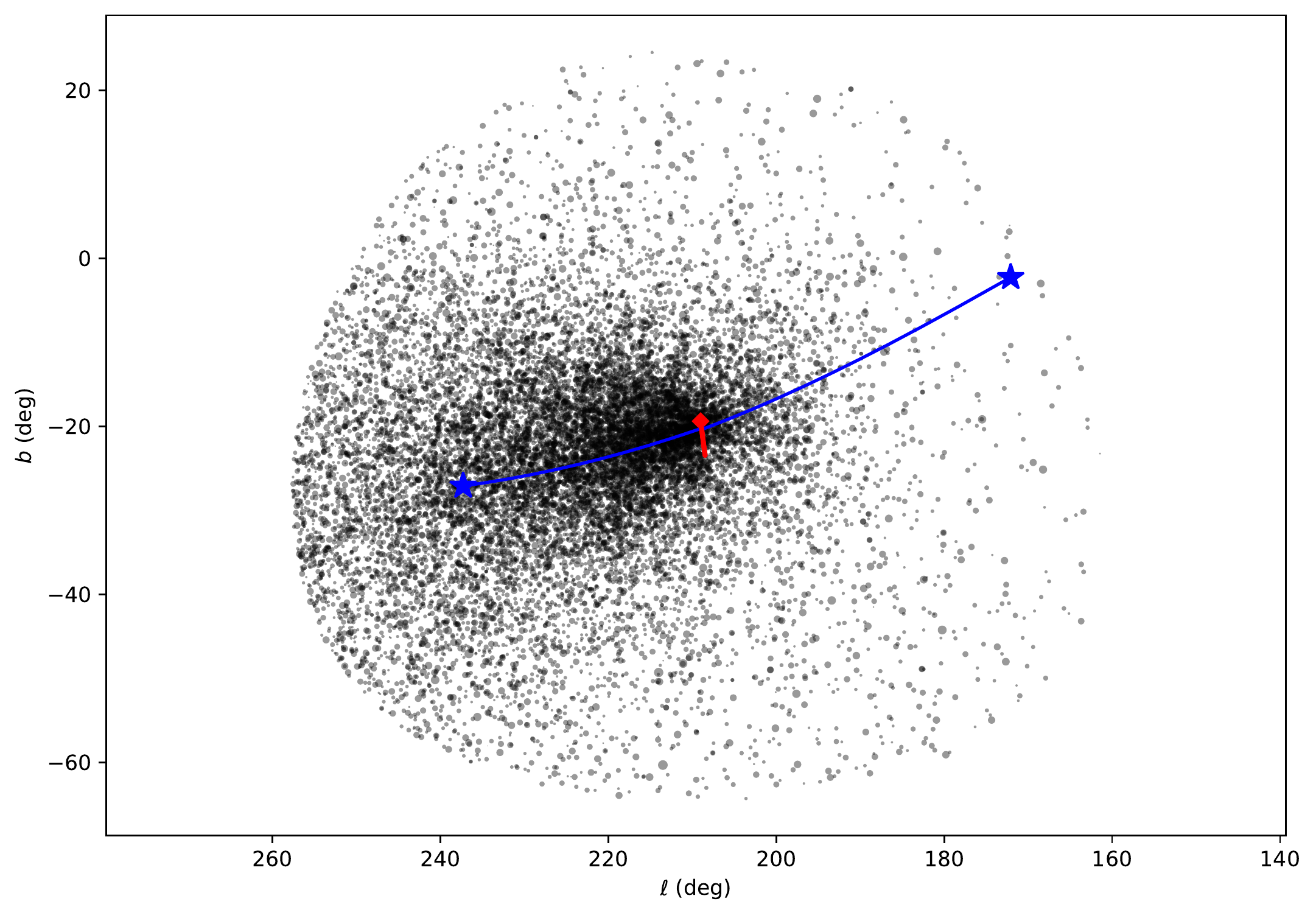}
    \caption{Positions of sources selected using the 2D traceback method described
            in \S\ref{sec:sample}. The sizes of
    the points are proportional to G band luminosity. The red diamond shows the position of the ONC and
            the red line shows its trajectory during the last 5 Myr. Blue stars
    shows the positions of \aeaur\ and \mucol\ and the blue lines show their trajectories.    }
    \label{fig:skyplot}
\end{figure*}

\begin{figure*}
        \centering
        \includegraphics[width=\textwidth]{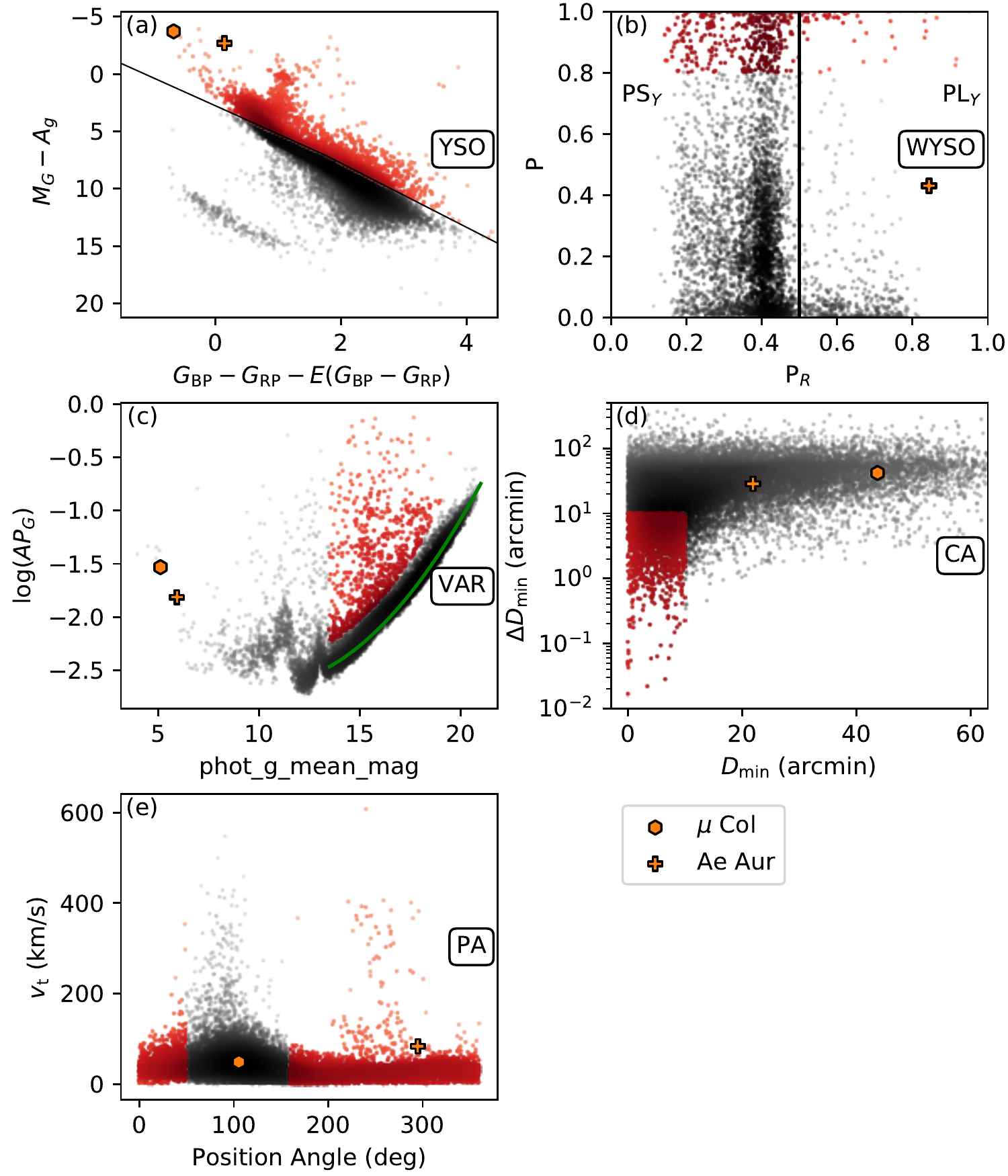}
        \caption{Five tests of youth and kinematic properties to flag candidates
                that have a higher chance of being actual ejected members of the
                ONC. In each panel, grey and black dots show the main 2D traceback
                sample of $\sim17,000$ sources, while red dots highlight the
                selected sources based on each criterion. The two orange
                symbols show the positions of \mucol\ and \aeaur. Each panel shows one
                criterion on which a source is tested to see if it will be flagged.
                \textbf{(a)} YSO flag, for sources fulfilling the optical
                color-magnitude cut (see text) in order to clean the sample of
                low-mass main sequence stars. \textbf{(b)} WYSO flag, highlighting
                sources that have a high probability of being a YSO based on IR
                colors (see text).  On the left side of the vertical line P$_R =
                0.5$ the PS$_Y$ value is plotted, while on the right side of the
                line PL$_Y$ is plotted (see text). \textbf{(c)} Variability flag
                (PV) using variability of \gaia-observed G band magnitude as a
                signature of variability (see text). \textbf{(d)} Closest approach
                flag (CA) by which the best astrometric candidates are selected
                (see text). \textbf{(e)} Position Angle flag (PA), used to select
                sources away from the zone contaminated most severely by galactic
                streaming.  }
        \label{fig:flags}
\end{figure*}

Figure~\ref{fig:skyplot} shows the overall distribution of the $\sim$17,000 sources
that satisfy the 2D traceback condition, described in \S\ref{sec:sample}. Given the
number of sources detected by this method and their asymmetric distribution in
position angle (\positionAngle) around the ONC, it is clear that the large majority are
contaminants from Galactic field stars that have apparent past positional overlap
with the ONC. It is expected that this occurs especially due to systematic
differential rotation in Galactic orbits, which we will refer to as ``Galactic
streaming'', thus explaining the asymmetry of the distribution, preferentially in
one direction from the ONC that is parallel to the Galactic plane.

\section{Sample Refinement from Signatures of Youth and Kinematic Properties}%
\label{sec:youth}

Given the large number of sources found by 2D traceback, the next challenge is thus
to find ways to filter out most of these contaminants to identify candidates that
have a higher likelihood of being real runaways. These could be targets for
spectroscopic follow-up, e.g., for radial velocity measurement and better stellar
characterization. To do this filtering, we now carry out six further tests,
focusing on aspects of stellar youth, via: 1) optical colors; 2) IR excess colors;
3) variability; and kinematic properties, via: 4) accuracy of coincidence with the
ONC centre; 5) if PA is away from the main contamination zone due to Galactic
streaming; 6) radial velocity considerations to achieve 3D traceback. 

If a star passes a test, we say it is ``flagged'' as being of greater potential
interest. However, some stars are not able to be examined for all the tests, and so
we will generally pass or ``flag'' such a star in respect of that test to err on
the side of inclusion. To distinguish such cases we use a numerical value for the
three states of a source with respect to a given flag: pass the flag (+1); fails
(-1); and could not be tested (0). 

\subsection{Youth via optical color-magnitude (YSO flag)}\label{sec:ysoflag}

We expect most lower-mass stars formed in the ONC to be in the pre-main sequence
phase, which may be assessed from the HR diagram. In particular, we examine the
position in the color-magnitude diagram (corrected by extinction and reddening) to
remove contaminating lower-mass main-sequence stars, following conditions used
previously by \cite{Kounkel2018,McBride2019}, i.e., the following cuts in
colour-magnitude space in the \gaia\ color-magnitude system (see
Fig.~\ref{fig:flags}a):
\begin{eqnarray}
M_{\rm G} < 2.46 \times |G_{\rm BP} - G_{\rm RP}| + 2.76 &;
        & |G_{\rm BP} - G_{\rm RP}| < 1.8 \nonumber \\
M_{\rm G} < 2.8 \times |G_{\rm BP} - G_{\rm RP}| + 2.16 &;
        & |G_{\rm BP} - G_{\rm RP}| \geq 1.8. \nonumber 
\end{eqnarray}
The extinction correction in this analysis was achieved following the method of
\cite{Zari2018} for the studied region. It consists of making a 3D grid on the
studied region and using the values for $G$ band extinction, $A_G$, and color
excess $E(G_{\rm BP} - G_{\rm RP})$ provided by \gaia. While the individual values
for extinction and color excess are not especially accurate, it is possible to use
an average for sources in each bin \citep{Zari2018,Andrae2018}.  We download a
special sample for this purpose from the same region of our original sample, using
the following conditions \citep{Zari2018}:
\begin{itemize}
    \item $M_{G} \leq 4.4 $
    \item $(G_{\rm BP} - G_{\rm RP} )\leq 1.7$ mag
    \item $\varpi/\sigma_{\varpi} > 5$
\end{itemize}
We then grid the region in 3D using bins of 10\,pc, and take the average on each
bin for extinction and color excess, obtaining a 3D extinction map. 

Using this method we found that 2,893 sources out of the $\sim$17,000 2D traceback
main sample have properties consistent with YSOs (or higher-mass main sequence
stars), of which only 10 sources have \texttt{phot\_bp\_rp\_excess\_factor}
higher than 2 (and these are below 2.5).

\subsection{\allwise\ IR classification (WYSO flag)}\label{sec:wysoflag}

The recent study by \cite{Marton2019a} performed a probabilistic classification of
YSOs in the \gaia\ catalogue using the cross-matched table between \gaia\ and the
\allwise\ database by \cite{Marrese2019}. The \allwise\ source catalogue
\citep{Cutri2013} is an extension of the \emph{Wide-field Infrared Survey Explorer
(WISE)} survey \citep{Wright2010} that contains 747 million sources with accurate
infrared photometry.  \emph{WISE} scanned the whole sky using four near-infrared
bands at 3.4, 4.6, 12 and 23\,$\mu$m, hereafter W1, W2, W3 and W4, respectively. 

\cite{Marton2019a} used a Machine Learning approach to classify sources in four
categories, i.e., Main Sequence stars (MS), Extragalactic objects (EG), Evolved
stars (E) and Young Stellar Objects (YSO). For the YSO classification, they used as
training sample a collection of photometric and spectroscopic YSO catalogues listed
by VizieR \citep[see Appendix A, B and C in][for references]{Marton2019a}, Spitzer
YSOs and YSO candidates from \citep{EvansII2003}. Applying a Random Forest
classification, they were able to recover 93.9\% of the training set correctly.
Using this method they provided a probability, PL$_{\rm Y}$, of a source being a
YSO using the W1-4 bands (among other features).

However, most \allwise\ sources had spurious photometry and point source
identification in the longest bands (W3 and W4) that could lead to false
classification. Therefore the whole classification was done also using only the W1
and W2 bands, providing a complementary probability, PS$_{\rm Y}$, of being a YSO
when discarding W3 and W4. Following \cite{Koenig2014} they also used random forest
classification to characterize with a probability, P$_{\rm R}$, for W3 and W4 to be
real.

Following their method, we used the \cite{Marton2019a} catalog and flagged sources
as Wise Young Stellar Object (WYSO) stars where: 
\begin{eqnarray}
\label{eq:wysoflag}
{\rm PL_Y} \geq 0.8 & \text{if} & {\rm P_R} > 0.5  \nonumber\\
{\rm PS_Y} \geq 0.8 & \text{if} & {\rm P_R} \leq 0.5.
\end{eqnarray}

Unfortunately, only  4183 sources, i.e., 25\%, 2D traceback sample have \allwise\
photometry. Of these, we find 420 sources that fulfil the WYSO criteria (see
Fig.~\ref{fig:flags}b). We choose not to penalize sources that could not be
evaluated using this method.  This essentially means that we flag WYSO = 0 any
source that was not in the \cite{Marton2019a} catalog. We use WYSO = +1 for
 sources that do have \allwise\ photometry and that fulfill Eq.~\ref{eq:wysoflag}.

\subsection{Variability (VAR flag)}\label{sec:pvflag}

The majority of YSOs exhibit variability \citep[e.g.,][]{Cody2014}. Thus, we expect
most true ONC members will do so also.  In \gaia\ DR2, only average values for the
photometry are published for all sources, although each source has been observed at
several epochs \citep{Evans2018}.  The reported value of the mean flux has an
associated uncertainty related to it.  Variability is assessed as proportional to
the standard deviation of the magnitude measurements, which can be reconstructed
from the mean quantity of the flux, the flux error and the number of measurements
for a given source.  In \gaia, the \gband\ band is the most precise photometric
measurement, thus we use it to construct a proxy for the amplitude of the variation
\citep{Eyer2019}. As amplitude proxy on \gband, hereafter $AP_\gband$, we use its
fractional standard error, i.e., $AP_\gband = \sigma(\gflux) / \langle \gflux
\rangle$. This value is obtained from \gaia\ DR2 as:
\begin{eqnarray}
        AP_\gband & = & \frac{\sigma(\gflux) } {\langle \gflux \rangle} \nonumber\\
     & = & \frac{ \sqrt{ \texttt{phot\_g\_n\_obs} } }
                         { \texttt{phot\_g\_mean\_flux\_over\_error} },  
\end{eqnarray}
where \texttt{phot\_g\_n\_obs} is the number of observations used to construct
$\langle \gflux \rangle$ and \texttt{phot\_g\_mean\_flux\_over\_error} is $\langle 
\gflux \rangle$ divided by its error divided by $\sqrt{ \texttt{phot\_g\_n\_obs} }$.

In Fig.~\ref{fig:flags}c we plot $AP_\gband$ against $\langle \gflux \rangle$.  A noise
threshold function ($PA_0$) is fit to the densest area of Figure~\ref{fig:flags}c,
shown with the green line, and having the form:
\begin{eqnarray}
\log[f(\gband)]  &=&  \frac{( \langle \gband \rangle - 10)^{2.04}} {70} - 2.65 ; 
                 \langle \gband \rangle > 13.5.
\end{eqnarray}
We quantify intrinsic variability ($V_i$) via:
\begin{eqnarray}
V_i &=& \log(AP_\gband)^2 - \log[f(\gband)]^2,
\end{eqnarray}
This method is only well behaved for faint sources, since brighter sources suffer
from other sources of photometric errors that are not well described as random
noise. Thus, we only evaluate intrinsic variability for sources with
\texttt{phot\_g\_mean\_mag}~$ = \langle \gband \rangle >13.5$.

Finally, we flag as VAR = +1 all sources with $\langle \gband\rangle > 13.5$ and
$V_i>1$. Sources with $\langle \gband\rangle < 13.5$, which could not be evaluated using
this method, are not penalized and are given a VAR = 0 flag. Overall 2,494 sources
are flagged VAR $\ge$ 0, of which 676 are flagged VAR = 1.

\subsection{Close Approach (CA flag)}\label{sec:CAflag}

Even with the unprecedented improvement in proper motion accuracy provided by
\gaia, many stars still have relatively large uncertainties in their astrometric
solutions. For stars that are further from the ONC, these uncertainties, plus those
associated with the effects of the Galactic potential that we have not accounted
for, have a correspondingly larger effect on the predicted position when the star
was near the ONC. So far we have been quite generous in the closest approach
distance to the ONC that is needed to select stars, i.e., it could be as large as
68.5\arcmin\ ($\sim 8$\,pc) for sources that are currently 45\deg\ away from the
ONC, with this limit set to be able to recover \mucol\ and \aeaur\ (see above).

Now, we wish to flag those sources that do have more accurate estimates of their
proper motions that bring them within 10\arcmin\ of the ONC centre and with an
uncertainty smaller than 10\arcmin. Most runaways are expected to be
ejected from the dense central region of the ONC, so with this flagged subset we
expect to have a higher likelihood of finding real runaways and reduce the level of
contamination compared to the main sample. The sources selected by this method are
shown in Figure~\ref{fig:flags}d: there are 1,447 sources out of the main sample of
$\sim17,000$. Thus we can see that even with this more restricted 2D traceback
condition, the sample is still likely to be dominated by contaminants. We also
remind that there could be true runaways, especially more distant ones like
\mucol\ and \aeaur, that are not selected by this method.

\subsection{Position Angle (PA flag)}\label{sec:PAflag}

A large degree of contamination is present after the 2D traceback selection, due to
streaming in the Galactic plane. This is evident from the asymmetric distribution
of the sources around the ONC (see Fig.~\ref{fig:skyplot}). We have thus added a
flag based on the position angle (\positionAngle) of a star's current angular position relative
to the ONC, where 0\deg\ is in the direction of the Galactic north pole. We flag
sources that are outside the range $50\deg<\positionAngle<160\deg$  with PA flag = 1,
which is the main contamination zone. Sources within such \positionAngle\ range are
flagged with PA = -1. Note that PA refers to the flag, while \positionAngle\
refers to the actual position angle in degrees.

\subsection{Radial Velocity Flag (RV flag)}
\label{sec:rv}

\begin{figure}
    \centering
    \includegraphics[width=\columnwidth]{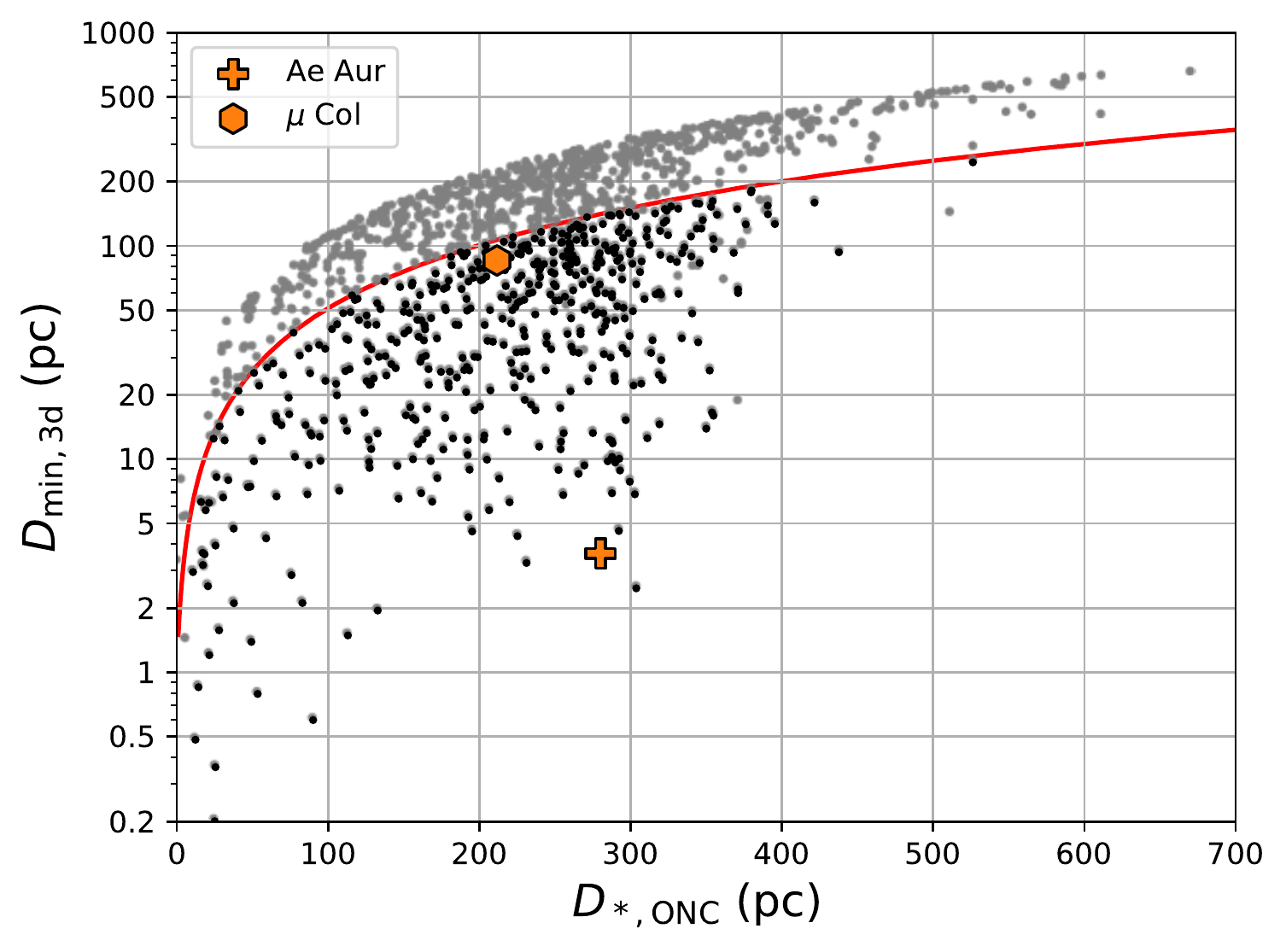}
    \caption{
        For those sources with measured radial velocities, the closest approach
        distance of the 3D trajectory to the ONC center, $D_{\rm min,3D}$, is shown
        versus the current 3D distance of the star from the ONC, $D_{\rm *,ONC}$.
        We select those sources that satisfy the threshold condition of
        Eq.~(\ref{eq:rvsel}), shown by the red line. The two orange symbols are
        \aeaur\ and \mucol, as labelled.
    }
    \label{fig:rvsel}
\end{figure}

\begin{figure*}
    \centering
    \includegraphics[width=\textwidth]{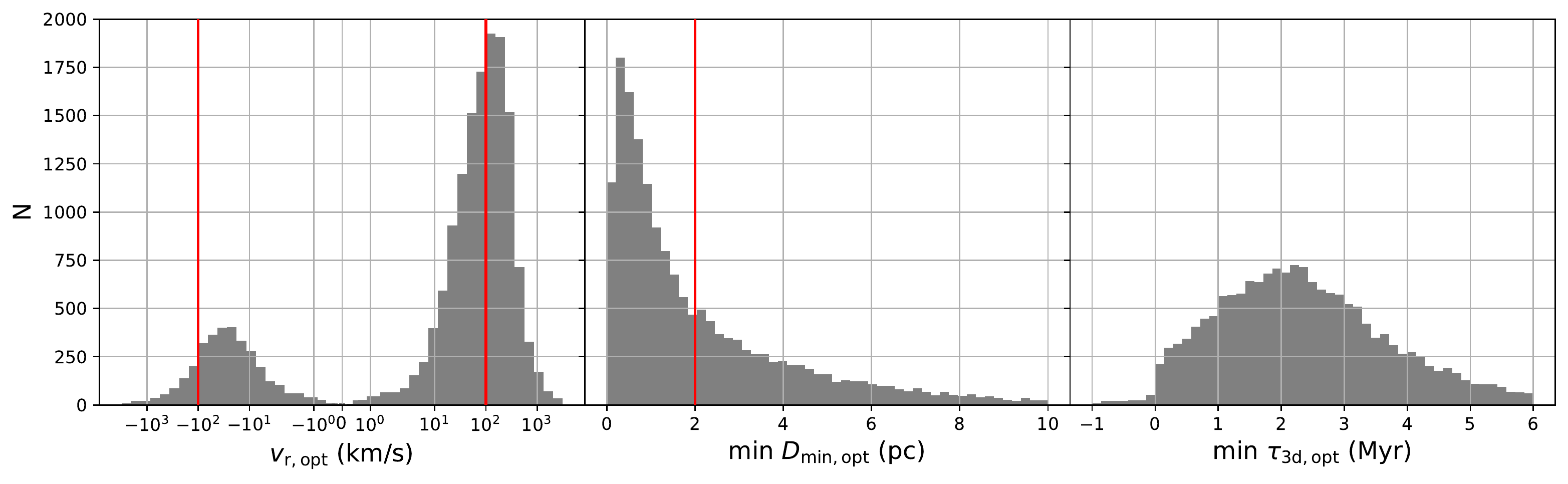}
    \caption{Results of the minimization of $D_{\rm min,3d}(v_{\rm r})$ for the
    optimal radial velocity $v_{\rm r,opt}$ (left panel), the minimum possible
    closest approach $D_{\rm min,opt}$ (middle panel) and minimum possible trace
    back time $\tau_{\rm 3d, opt}$ (right panel). Red vertical lines shows the
    limits discussed in the text.} 
    \label{fig:vropt}
\end{figure*}

\subsubsection{Stars with measured radial velocities}

In the 2D traceback sample of $\sim17,000$ sources, only about 7\% have measured
radial velocities, i.e., $1,162$ stars. However, for these sources, we are able to
carry out a 3D traceback analysis, which is more restrictive than the 2D method.

Using simple vector algebra, we calculate the closest approach to the ONC
in 3D, $D_{\rm min,3D}$. Given the position $\vec{X_*}$ and velocity
$\vec{V_*}$ of a star and the position $\vec{X_{O}}$ and velocity $\vec{V_{O}}$
of the origin, in this case the ONC, the time of the closest approach to $O$ is
\begin{eqnarray}
\label{eq:tmin3d}
\tau_{\rm min,3D} &=& - \frac{(\vec{X_*} - \vec{X_O})\cdot(\vec{V_*} -
\vec{V_O})}{  |\vec{V_*} - \vec{V_O}|^2   }.
\end{eqnarray}
Then, the closest approach distance is:
\begin{eqnarray}
\label{eq:dmin3d}
D_{\rm min,3D} &=& |(\vec{X_*} - \tau_{\rm min,3D}\vec{V_*}) - (\vec{X_O} -
\tau_{\rm min,3D}\vec{V_O})|.
\end{eqnarray}

Following a similar approach to that used in the 2D traceback described in
\S\ref{sec:sample}, we adopt a closest approach threshold that grows as a function
of current distance of the source from the ONC:
\begin{equation}
    D_{\rm min,3D} / {\rm pc} < 1 + 0.5 (d_{\rm ONC} / {\rm pc})
    \label{eq:rvsel}
\end{equation}
where $d_{\rm ONC}$ is the current 3D distance of a star to the ONC.  The form of
this equation was guided by consideration of \aeaur\ as can be seen in 
Figure~\ref{fig:rvsel}, where we illustrate this threshold.
While \aeaur\ and \mucol\ do not have measured radial velocities with \gaia, we
made use of the values from Hipparcos used in the analysis of
\citet{Hoogerwerf2001}.  With these central values for radial velocity and distance
for \aeaur, the past trajectory misses the ONC by $\sim$100\,pc. However, we note that
there are significant spreads in the distributions of the properties of the stars
used by \cite{Hoogerwerf2001}, while the affect of accelerations induced by the
Galactic potential and Orion's giant molecular clouds' potentials will also impart
apparent discrepancies.

Using the criterion described in Eq.~(\ref{eq:rvsel}) selects 516 sources as
ejection candidates out of the 1200 sources with radial velocities.  However, for a
small number, 25, of these selected sources the obtained $\tau_{\rm
min,3D}$ is negative, i.e., their closest approach in 3D to the ONC is in the
future. We discard such sources, leaving a final selected sample of 491 out of
1,200, i.e., 40\%.

\begin{table*}
    \centering
    \tabletypesize{\sriptsize}
    \begin{tabular}{cccccc|l}
     YSO & WYSO & VAR & CA & PA & RV &Flag Combinations\\\hline
 2893 & 420(13231) & 676(2494) & 1447 & 4755 & 6572 &\\\hline
   & 205(1989) & 266(1588) & 464 & 906 & 833&YSO\\
   &    & 150(1923) & 191(894) & 151(2963) & 269(5280)&WYSO\\
   &    &    & 324(547) & 260(812) & 397(1076)&VAR\\
   &    &    &    & 540 & 480&CA\\
   &    &    &    &    & 2109&PA\\
\hline   &    & 104(1198) & 97(278) & 63(498) & 146(637)&YSO $\times$ WYSO\\
   &    &    & 149(307) & 101(495) & 207(650)&YSO $\times$ VAR\\
   &    &    &    & 191 & 192&YSO $\times$ CA\\
   &    &    &    &    & 308&YSO $\times$ PA\\
   &    &    & 95(425) & 45(555) & 133(851)&WYSO $\times$ VAR\\
   &    &    &    & 52(290) & 110(355)&WYSO $\times$ CA\\
   &    &    &    &    & 103(1446)&WYSO $\times$ PA\\
   &    &    &    & 138(233) & 213(290)&VAR $\times$ CA\\
   &    &    &    &    & 154(371)&VAR $\times$ PA\\
   &    &    &    &    & 193&CA $\times$ PA\\
\hline   &    &    & 60(224) & 32(328) & 95(495)&YSO $\times$ WYSO $\times$ VAR\\
   &    &    &    & 29(111) & 66(154)&YSO $\times$ WYSO $\times$ CA\\
   &    &    &    &    & 50(212)&YSO $\times$ WYSO $\times$ PA\\
   &    &    &    & 64(138) & 120(181)&YSO $\times$ VAR $\times$ CA\\
   &    &    &    &    & 82(237)&YSO $\times$ VAR $\times$ PA\\
   &    &    &    &    & 90&YSO $\times$ CA $\times$ PA\\
   &    &    &    & 31(178) & 79(246)&WYSO $\times$ VAR $\times$ CA\\
   &    &    &    &    & 42(264)&WYSO $\times$ VAR $\times$ PA\\
   &    &    &    &    & 36(130)&WYSO $\times$ CA $\times$ PA\\
   &    &    &    &    & 93(128)&VAR $\times$ CA $\times$ PA\\
\hline   &    &    &    & 21(99) & 52(146)&YSO $\times$ WYSO $\times$ VAR $\times$ CA\\
   &    &    &    &    & 31(164)&YSO $\times$ WYSO $\times$ VAR $\times$ PA\\
   &    &    &    &    & 22(69)&YSO $\times$ WYSO $\times$ CA $\times$ PA\\
   &    &    &    &    & 54(86)&YSO $\times$ VAR $\times$ CA $\times$ PA\\
   &    &    &    &    & 28(107)&WYSO $\times$ VAR $\times$ CA $\times$ PA\\
\hline   &    &    &    &    & 20(67)&YSO $\times$ WYSO $\times$ VAR $\times$ CA $\times$ PA\\
\hline
    \end{tabular}
    \caption{Number of Sources that fulfill all combination of flags described
    in the text. Quantities between parenthesis show the number of sources that
    fulfill the condition and also have all the information required (see
    text). The table is organized on blocks where, from top to bottom, results
    show the combination of 1, 2, 3, 4, 5 and 6 flags respectively.}
    \label{tab:flags}
\end{table*}

\begin{table}
\centering
\begin{tabular}{lc|ccc}
     Score        &   Subclass           &YSO&    WYSO              &     VAR              \\ \hline
\multirow{3}{*}{a}&     I             & 1 &     1                &      1               \\ \cline{2-5}
                  &     II            & 1 &$_\text{ 0}^\text{ 1}$&$_\text{ 1}^\text{ 0}$\\ \cline{2-5}
                  &     III           & 1 &     0                &      0               \\ \hline
\multirow{5}{*}{b}&\multirow{2}{*}{I} &-1 &     1                &      1               \\
                  &                   & 1 &$_\text{-1}^\text{ 1}$&$_\text{ 1}^\text{-1}$\\ \cline{2-5}
                  &\multirow{2}{*}{II}&-1 &$_\text{ 0}^\text{ 1}$&$_\text{ 1}^\text{ 0}$\\
                  &                   & 1 &$_\text{ 0}^\text{-1}$&$_\text{-1}^\text{ 0}$\\ \cline{2-5}
                  &       III         &-1 &     0                &      0               \\ \hline
\multirow{3}{*}{c}&\multirow{2}{*}{I} &-1 &$_\text{-1}^\text{~1}$&$_\text{ 1}^\text{-1}$\\
                  &                   & 1 &    -1                &     -1               \\ \cline{2-5}
                  &           II      &-1 &$_\text{ 0}^\text{-1}$&$_\text{-1}^\text{ 0}$\\ \hline
                d &           I       &-1 &    -1                &     -1               \\ \hline
\end{tabular}
\caption{
        Summary table of score system used to classify and label candidate runaway
        sources. Score is based in how many signatures of youth a source fails
        (indicated with -1), while the subclass column is based on how many flags
        for which the source could not be tested (indicated with 0). }
\label{tab:score}
\end{table}

\subsubsection{Required radial velocity}
\label{sec:orvflag}

In the 2D traceback selected sample 93\% of sources do not have radial velocities.
This is the final parameter from the 6-dimensional space needed to fully
characterize the trajectory of a star.  For the sources lacking radial velocity
measurements, we have calculated the value of radial velocity, $v_{\rm r,opt}$
needed so that their past trajectory has the closest approach to the ONC center.
To account for measurement errors, we carry out Monte Carlo sampling over the
distributions of astrometric parameters, assuming Gaussian distributions for the
uncertainties, to obtain not only a distribution of $v_{\rm r,opt}$, but also of
closest approach distances and traceback times. From these distributions we report
the 16 and 84 percentiles for each source.

For sampling on $\varpi$, simple Gaussian sampling is not enough
\citep{Bailer-Jones2015}. Even though we have chosen to work with sources with
small enough errors so we can use distances as $r = 1/ \varpi$, sampling distances
using a Gaussian distribution around $\varpi$ caused us to lose some sampling
probability when $\varpi < 0$. Instead, to infer the distribution of
distances that a source with a given $\varpi$ and $\sigma_\varpi$ would have, we
use the posterior used in \cite{Bailer-Jones2018} assuming an exponentially
decreasing space density prior \citep[see][for further
details]{Bailer-Jones2015,Bailer-Jones2018}.

Figure~\ref{fig:vropt} shows distributions of $v_{\rm r,opt}$ and the minimum
values (16th percentile) of the distributions for $D_{\rm min,opt}$ and $\tau_{\rm
3D,opt}$. The first panel shows that the majority of sources need positive radial
velocities to reach the ONC during their past trajectories, since most of them are
at distances $>$410\,pc. It is important to note that $v_{\rm r,opt}$ generally has
large uncertainties. Most of this comes from parallax uncertainties, which can give
a large range of possible distances, directly affecting the radial
velocity needed to reach the ONC. 

The second and third panels in Figure~\ref{fig:vropt} show the 16th percentile of
the distributions of $D_{\rm min,opt}$ and $\tau_{\rm 3d,opt}$. An important point
to note is that tracing back in 3D will not necessarily give the same result as 2D
traceback. There are two reasons. First, in 2D traceback we assume the proper
motion is constant along the trajectory, which is an approximation that becomes
less valid for sources with relatively large current angular separations from the
ONC.  Second, the 2D traceback method does not consider the radial velocity of the
ONC.  For these reasons, the traceback time in 3D can be different from that in 2D.

A difference in the traceback time affects the final position in the sky of closest
possible approach to the ONC.  If the ONC radial velocity were zero, then the
minimum closest approach from a source to the ONC would be given by the 2D closest
approach in the plane of the sky. However since it moves, the closest distance may
be different from its 2D counterpart. The result is that the best 3D closest
approach of some sources is larger than the threshold used in the 2D traceback. We
therefore exclude those sources where the minimum (16th percentile) closest
approach does not satisfy the 3D traceback threshold.

Conditions on $v_{\rm r,opt}$, $D_{\rm min,opt}$ and $\tau_{\rm 3D,opt}$ can thus
be used as thresholds to exclude some sources from the candidate list.  First, we
consider the magnitude of $v_{\rm opt}$. We do not expect that dynamically ejected
stars are likely to have radial velocities greater than 1000\,\kms\ as the maximum
ejection speed is approximately the escape speed from the location of ejection,
which is limited by the escape speed from near the surface of the ejecting star.
Indeed, known runaway stars with velocities $>100\,\kms$ are very rare. If this
velocity were to be used as a threshold, then 7,231 sources would be excluded.  An
additional 3,267 sources  can be discarded if we exclude any source has min $D_{\rm
min,opt}>2$\,pc, i.e., about twice the ONC half-mass radius.  In principle, the
revised traceback time could also be used to discard sources, however the range of
values shown by $\tau_{\rm 3D,opt}$ are similar to those found earlier in the 2D
traceback method, so we do not exclude any based on this quantity.

In summary, by assessing the conditions needed for 3D traceback in the sample of
sources that do not have radial velocity measurements, we have excluded 6,496
sources out of \Ntraceback. Thus every source is given an entry for the RV flag,
but with the selection criteria depending on whether it is a source with a measured
radial velocity or not.

\section{Results}\label{sec:results}

We now discuss the results of applying the sample refinement criteria described in
the last section.

\subsection{Flag combinations}\label{sec:candidates}

Table~\ref{tab:flags} shows the number of stars that satisfy the various possible
combinations of flags. For the cases of WYSO and VAR flags, not all sources could
be tested, either because the sources did not appear in the \allwise\ catalog (see
\S\ref{sec:wysoflag}) or were too bright (see \S\ref{sec:pvflag}). As mentioned in
\S\ref{sec:youth}, we use a numerical value for each flag when passing (+1),
failing (-1) or when it could not be tested (0). Effectively only WYSO and YSO
flags have values equal to zero. The numbers in Table \ref{tab:flags} show the
combination of positive cases, while numbers in parentheses show the combinations
of flags with +1 or 0 values.

In order to select the best candidates for high velocity runaways, we have
developed a score system which gives an unique score to each source based on the
number of flags that it meets and to the transverse velocity it has within the
context of each group. There are four general scores given, as described below. 

Sources that do not fail any young signature flag (i.e., YSO, WYSO and VAR) are
given a score \emph{a}. Sources that fail one signature of youth flag are scored
with \emph{b}.  Sources that fail two signatures of youth are scored as \emph{c} and
sources that fail all three YSO, WYSO and VAR flags are scored as \emph{d}.  Within each
score we added three subclasses depending on how many zeros are found in the
signatures of youth flags.  Subclass I means there is no zero in these flags,
subclass II means there is one of the flags with a zero value, while subclass III
means there are two flags with zero values.  Therefore we have general scores as
aI, aII, aIII, bI, bII, bIII, cI, cII and d. Table \ref{tab:score}
summarizes this scoring system. 

Three modifiers are added to the score label depending on the astrometric flags CA,
PA and RV. A ``+" character is added if a source passes the CA flag, i.e., is a
particularly good candidate whose trajectory overlaps within
10\arcmin$\pm$10\arcmin\ with the ONC. A ``*" character is given to sources that
fail the PA flag. This reminds us of sources that are more likely to be
contaminants due to Galactic streaming. Finally, a ``!" character is added for
sources that fail the RV flag, i.e., sources that are unlikely to come from the
center of the ONC given their current astrometric parameters and uncertainties.

Finally, within each group of scores, e.g. aI+*!, an identifier is appended
depending on its estimated transverse velocity on the frame of reference of the
ONC, $v_t'$. Where, from the fastest to the slowest candidate, a sorted ordinal
number is used as identifier. Since each score is unique, we will also use it as a
label for each source in the catalog. For instance, the best candidate found in our
catalog is source aI+1 corresponding to the source DR2 3209590577396377856 (see
Table~\ref{tab:bestcandidates}). Another strong candidate, except that it happens
to be in the contaminated zone, is source aI+*1 which corresponds to source DR2
3015321754828860928.

\subsection{Already known ONC members}%
\label{sec:known}

Before presenting the selection of sources based on the flag system, we first
discuss how the sample of known ONC members are represented in our selection
system.  This is motivated by the fact that from the main sample of 2D traceback
selected stars, i.e., totalling \Ntraceback\ sources, there are 67 \emph{a+} sources
(i.e., that do not fail any flags), of which 20 are scored \emph{aI+} (i.e., pass
all 6 flags), and of these samples of 67 (20) sources, 54 (17) are already known
members of the Orion A complex.
\begin{longrotatetable}
\begin{deluxetable*}{rrrrrrrrrrrrrrr}
\tabletypesize{\scriptsize}
\tablecaption{%
Traceback candidates that are known members of the Orion A complex from
\citep{McBride2019} with transverse velocities above 4\,\kms. First column shows
the \gaia\ ID with alternative names for specific sources within parenthesis, such
as \aeaur, \mucol. The first section of the table shows candidate runaways reported
recently by \cite{McBride2019} with the label letters used by the authors for quick
reference. Second column shows the corresponding score and label system used in
this work. Unless otherwise stated, all astrometric parameters and others derived
from astrometry are taken from \gaia\ DR2 catalog.  Three dimensional distance
$d_{\rm ONC}$ is calculated using $1/\varpi$, and we also show the current angular
distance to the ONC, $\theta_{\rm ONC}$. We show the transverse velocity in the
frame of reference of the ONC, $v_{\rm t}'$ where we used the factors $\cos(\ell -
\ell_{\rm ONC})$ and $\cos(b - b_{\rm ONC})$ to subtract the correct velocity
components.  $t_{\rm back}$ is the 2D traceback time to the ONC (see
\S\ref{sec:sample}). We also show the values of parameters used in the selection
criteria, where values in bold show the cases where a criterion is fulfilled (see
\S\ref{sec:youth}). The 16th, 50th and 84th percentile of the required radial
velocity is shown by $v_{\rm r, opt}$, next to the actual \gaia\ DR2 radial
velocity measurement, $v_{\rm r}$, when available; both radial velocities were used
for the RV criterion.  Sources that pass the \emph{YSO} flag are marked with a
$\checkmark$ symbol in the YSO column. WYSO flag is based on \cite{Marton2019a} YSO
probabilities, P(YSO). A ``*" mark is used to indicate when P$_{\rm R}\leq0.5$ and
PL$_{\rm Y}$ was used, otherwise PS$_{\rm Y}$ is shown. The calculated intrinsic
variability used by the VAR flag, estimated in \S\ref{sec:pvflag} is shown in the
$IV$ column. The CA flag is based on the closest 2D traceback angular distance to
the ONC, \dmin. Finally, the position angle, \emph{\positionAngle}, with respect to
the ONC is shown in the last column, where bold symbols show sources in the
``clean" zone.%
\label{tab:known}}
\tablehead{
\colhead{Gaia ID}  &\colhead{Label}  &\colhead{$\ell\,(^\circ)$}  &\colhead{$b\,(^{\circ})$}  &\colhead{$d_{\rm ONC}$}  &\colhead{$\theta_{\rm ONC}$}  &\colhead{$v_{t}'$}  &\colhead{$t_{\rm back}$}  &\colhead{$v_{\rm r,opt}$}  &\colhead{$v_{\rm r}$}  &\colhead{YSO}  &\colhead{P(YSO)}  &\colhead{$V_i$}  &\colhead{$D_{\rm min}$}  &\colhead{$\theta_{\rm PA}$}  \\
\nocolhead{None} &\colhead{Score} &\colhead{[J2015.5]} &\colhead{[J2015.5]} &\colhead{(pc)} &\colhead{($^\circ$)} &\colhead{(km/s)} &\colhead{(Myr)} &\colhead{(km/s)} &\colhead{(km/s)} &\nocolhead{None} &\colhead{ } &\colhead{ } &\colhead{(')} &\colhead{($^\circ$)} \\
}
\startdata
  \textbf{(u)3017373645390119040} & aIII+!4 & 208.8720 & -19.2685 & 38.77 & 0.18 & 39.04 & 0.03 & $-3492^{+1226}_{-3178}$ &   & $\checkmark$ &   &   & \textbf{9.96}$\pm$\textbf{0.22} & \textbf{310.20} \\ 
 \textbf{(d)3209653627514662528} & bI+!4 & 207.9739 & -18.9000 & 34.13 & 1.14 & 26.90 & 0.30 & $ 161^{+44}_{-41}$ &   & $\checkmark$ & 0.40 & \textbf{2.25} & \textbf{1.61}$\pm$\textbf{2.56} & \textbf{295.07} \\ 
 \textbf{(h)3209624872711454976} & aI+2 & 208.2961 & -19.1976 & 16.83 & 0.73 & 18.56 & 0.25 & $\mathbf{  -6^{+24}_{-22}}$ &   & $\checkmark$ & \textbf{0.87$^*$} & \textbf{4.78} & \textbf{0.04}$\pm$\textbf{2.10} & \textbf{284.66} \\ 
 \textbf{(i)3209497088842680704} & aI+3 & 208.3825 & -19.8859 & 18.51 & 0.81 & 16.35 & 0.36 & $\mathbf{  -6^{+16}_{-15}}$ &   & $\checkmark$ & \textbf{0.99} & \textbf{4.25} & \textbf{4.14}$\pm$\textbf{3.10} & \textbf{231.29} \\ 
 \textbf{(f)3017166907140904320} & bI+*1 & 209.8430 & -19.6422 & 31.55 & 0.87 & 16.26 & 0.30 & $\mathbf{ -45^{+20}_{-19}}$ &   & $\checkmark$ & 0.77 & \textbf{1.67} & \textbf{1.15}$\pm$\textbf{2.59} & 107.20 \\ 
 \textbf{(b)3209424108758593408} & bII+9 & 208.9286 & -19.5733 & 21.29 & 0.21 & 16.23 & 0.10 & $\mathbf{-138^{+75}_{-78}}$ & \textbf{10.67$\pm$11.93} & $\checkmark$ & 0.80 &   & \textbf{1.01}$\pm$\textbf{0.86} & \textbf{202.93} \\ 
 \textbf{(g)3209531650444835840} & bII+!29 & 208.7476 & -19.3145 & 3.18 & 0.27 & 14.01 & 0.12 & $\mathbf{  55^{+49}_{-42}}$ & 17.60$\pm$4.17 & $\checkmark$ & 0.21 &   & \textbf{0.47}$\pm$\textbf{1.05} & \textbf{284.91} \\ 
 \textbf{(a)3209424108758593536} & bII+13 & 208.9254 & -19.5771 & 17.13 & 0.22 & 13.91 & 0.13 & $\mathbf{ -69^{+64}_{-70}}$ &   & $\checkmark$ & 0.71 &   & \textbf{1.02}$\pm$\textbf{1.07} & \textbf{203.32} \\ 
\hline
  \textbf{3017367155707574784} & bII+!3 & 208.8952 & -19.3620 & 198.04 & 0.11 & 119.80 & 0.00 & $-160583^{+902}_{-960}$ &   &   &   & \textbf{2.74} & \textbf{6.55}$\pm$\textbf{0.03} & \textbf{280.96} \\ 
 \textbf{(Ae Aur)182071570715713024} & bII1 & 172.0812 & -2.2591 & 280.18 & 41.84 & 83.56 & 3.30 & $\mathbf{  57^{+5}_{-5}}$ & \textbf{57.50$\pm$1.20}\tablenotemark{a} & $\checkmark$ & 0.43 &   & $21.88\pm28.54$ & \textbf{294.88} \\ 
 \textbf{3017166323025342336} & cII+*!65 & 209.8963 & -19.5461 & 186.80 & 0.89 & 72.26 & 0.04 & $-2146^{+44}_{-45}$ &   &   & 0.67$^*$ &   & \textbf{6.50}$\pm$\textbf{0.36} & 100.35 \\ 
 \textbf{3015731254191486464} & bI+*!6 & 212.5260 & -18.8654 & 343.59 & 3.52 & 49.26 & 0.08 & $-697^{+9}_{-9}$ &   &   & \textbf{0.88} & \textbf{1.61} & \textbf{7.51}$\pm$\textbf{0.65} & 81.61 \\ 
 \textbf{($\mu$ Col)2901155648586891648} & aIII*80 & 237.2863 & -27.1021 & 211.65 & 28.31 & 48.36 & 3.91 & $\mathbf{  61^{+16}_{-13}}$ & \textbf{109.00$\pm$2.50}\tablenotemark{a} & $\checkmark$ &   &   & $43.69\pm41.82$ & 105.27 \\ 
 \textbf{3017265794467742592} & cII+*!133 & 209.0880 & -19.5702 & 254.36 & 0.21 & 48.25 & 0.01 & $-10514^{+1327}_{-1632}$ &   &   &   & 0.90 & \textbf{7.55}$\pm$\textbf{0.09} & 156.91 \\ 
 \textbf{3017363131310582016} & bII+!10 & 209.0102 & -19.2647 & 269.67 & 0.12 & 45.62 & 0.01 & $-140779^{+481}_{-359}$ &   &   &   & \textbf{1.39} & \textbf{7.18}$\pm$\textbf{0.06} & \textbf{0.75} \\ 
 \textbf{3017246449935170304} & d+*!77 & 209.3468 & -19.4736 & 206.04 & 0.35 & 45.58 & 0.02 & $-3904^{+673}_{-960}$ &   &   & 0.73$^*$ & 1.00 & \textbf{0.18}$\pm$\textbf{0.23} & 104.85 \\ 
 \textbf{3017369217290106880} & aII+!2 & 208.9876 & -19.2373 & 118.14 & 0.16 & 31.86 & 0.04 & $5212^{+7664}_{-2650}$ &   & $\checkmark$ &   & \textbf{1.65} & \textbf{6.90}$\pm$\textbf{0.35} & \textbf{351.86} \\ 
 \textbf{3017266516022225920} & bII+*!53 & 209.0669 & -19.4662 & 58.29 & 0.10 & 24.34 & 0.02 & $-1823^{+452}_{-841}$ &   &   &   & \textbf{2.49} & \textbf{1.05}$\pm$\textbf{0.21} & 144.65 \\ 
 \textbf{3016867049702538368} & cI+*!78 & 210.7112 & -19.7078 & 112.38 & 1.71 & 23.44 & 0.33 & $-197^{+5}_{-6}$ &   &   & \textbf{0.82$^*$} & -0.29 & \textbf{5.44}$\pm$\textbf{2.76} & 100.77 \\ 
 \textbf{3017248094906815488} & cII+*!198 & 209.3396 & -19.4200 & 240.96 & 0.33 & 22.26 & 0.04 & $-1986^{+915}_{-3009}$ &   &   &   & 0.62 & \textbf{4.42}$\pm$\textbf{0.63} & 96.21 \\ 
 \textbf{3017266172424849664} & bII+*!56 & 209.0835 & -19.4899 & 298.41 & 0.13 & 21.09 & 0.01 & $-9476^{+2592}_{-6185}$ &   &   &   & \textbf{2.50} & \textbf{5.78}$\pm$\textbf{0.09} & 144.73 \\ 
 \textbf{3017242051888552704} & bII+*3 & 209.2926 & -19.7397 & 8.41 & 0.47 & 15.65 & 0.20 & $\mathbf{  19^{+23}_{-21}}$ &   & $\checkmark$ & 0.63$^*$ &   & \textbf{2.71}$\pm$\textbf{1.68} & 141.40 \\ 
 \textbf{3017240746218498176} & cI+*!86 & 209.3243 & -19.7989 & 90.63 & 0.53 & 15.41 & 0.32 & $ 483^{+134}_{-112}$ &   & $\checkmark$ & 0.23$^*$ & 0.46 & \textbf{9.20}$\pm$\textbf{2.83} & 142.73 \\ 
 \textbf{3015625563635553024} & bI+*2 & 211.7370 & -19.6999 & 31.49 & 2.71 & 14.25 & 1.05 & $\mathbf{  11^{+7}_{-6}}$ &   & $\checkmark$ & 0.73 & \textbf{2.58} & \textbf{4.02}$\pm$\textbf{9.01} & 96.60 \\ 
 \textbf{3017369217290105472} & bII+!28 & 208.9860 & -19.2301 & 177.85 & 0.16 & 14.25 & 0.11 & $-3038^{+1287}_{-3324}$ & 33.20$\pm$0.48 & $\checkmark$ & 0.30 &   & \textbf{9.34}$\pm$\textbf{1.01} & \textbf{351.62} \\ 
 \textbf{3017147592669974144} & cII+*!211 & 209.7571 & -19.7405 & 246.93 & 0.83 & 13.58 & 0.16 & $-619^{+20}_{-22}$ & 7.57$\pm$2.87 &   & 0.45 &   & \textbf{5.79}$\pm$\textbf{1.38} & 115.46 \\ 
 \textbf{3209520036854102912} & bII+!30 & 208.8184 & -19.5070 & 291.86 & 0.23 & 13.13 & 0.04 & $-3372^{+1273}_{-3975}$ &   &   &   & \textbf{1.98} & \textbf{9.51}$\pm$\textbf{0.41} & \textbf{237.10} \\ 
 \textbf{3209570000209415168} & d+!112 & 208.5551 & -18.6792 & 235.24 & 0.86 & 12.52 & 0.18 & $-448^{+37}_{-46}$ &   &   & 0.78$^*$ & 0.44 & \textbf{5.47}$\pm$\textbf{1.72} & \textbf{327.24} \\ 
 \textbf{3209519319595376512} & bII+!35 & 208.8678 & -19.4685 & 120.79 & 0.16 & 11.37 & 0.08 & $-3154^{+17090}_{-11742}$ & 16.18$\pm$2.42 & $\checkmark$ & 0.01$^*$ &   & \textbf{9.44}$\pm$\textbf{0.65} & \textbf{239.03} \\ 
 \textbf{3017341385903759744} & bI+*3 & 209.3795 & -19.3238 & 11.17 & 0.37 & 10.85 & 0.19 & $\mathbf{   9^{+34}_{-34}}$ &   & $\checkmark$ & 0.70$^*$ & \textbf{4.71} & \textbf{2.25}$\pm$\textbf{1.59} & 80.78 \\ 
 \textbf{3017270879709003520} & bII+15 & 208.9542 & -19.5277 & 13.25 & 0.16 & 10.20 & 0.14 & $\mathbf{ -35^{+65}_{-71}}$ &   & $\checkmark$ & 0.15$^*$ &   & \textbf{1.25}$\pm$\textbf{1.19} & \textbf{200.75} \\ 
 \textbf{3015714967674577024} & aI*1 & 211.1998 & -19.7645 & 21.49 & 2.20 & 10.12 & 1.29 & $\mathbf{  47^{+22}_{-18}}$ &   & $\checkmark$ & \textbf{1.00} & \textbf{2.91} & $4.00\pm12.31$ & 99.85 \\ 
 \textbf{3017359699643655552} & bIII+*!17 & 209.1160 & -19.3226 & 142.85 & 0.12 & 10.02 & 0.05 & $1948^{+806}_{-426}$ &   &   &   &   & \textbf{7.12}$\pm$\textbf{0.46} & 60.25 \\ 
 \textbf{3015334914608642688} & aI*2 & 212.5184 & -19.6520 & 30.79 & 3.48 & 9.79 & 2.09 & $\mathbf{  41^{+6}_{-5}}$ &   & $\checkmark$ & \textbf{0.90$^*$} & \textbf{2.61} & $14.10\pm18.37$ & 94.37 \\ 
 \textbf{3017251913133545984} & bII+*!63 & 209.2630 & -19.4002 & 26.65 & 0.25 & 9.77 & 0.16 & $ 289^{+388}_{-237}$ &   &   &   & \textbf{1.66} & \textbf{2.05}$\pm$\textbf{1.69} & 93.65 \\ 
 \textbf{3017260022031719040} & aII+1 & 209.1180 & -19.8014 & 15.04 & 0.45 & 9.67 & 0.36 & $\mathbf{   0^{+29}_{-27}}$ &   & $\checkmark$ &   & \textbf{3.76} & \textbf{3.70}$\pm$\textbf{3.11} & \textbf{165.32} \\ 
 \textbf{3209429503235627776} & cII+!81 & 208.7776 & -19.6471 & 183.18 & 0.36 & 9.23 & 0.19 & $-589^{+216}_{-680}$ &   &   &   & 0.64 & \textbf{4.61}$\pm$\textbf{2.54} & \textbf{221.29} \\ 
 \textbf{3017247240209309056} & cI+*!95 & 209.3667 & -19.4278 & 169.11 & 0.36 & 9.09 & 0.15 & $-725^{+59}_{-78}$ &   &   & \textbf{0.87$^*$} & -0.00 & \textbf{7.47}$\pm$\textbf{1.27} & 96.97 \\ 
 \textbf{3215872396561733504} & aII4 & 207.4842 & -18.8537 & 16.02 & 1.61 & 8.31 & 1.23 & $\mathbf{  23^{+4}_{-4}}$ & \textbf{20.63$\pm$3.20} & $\checkmark$ & \textbf{0.90} &   & $7.26\pm10.37$ & \textbf{289.18} \\ 
 \textbf{3017252600328207104} & aII+*!4 & 209.1973 & -19.5143 & 14.34 & 0.23 & 7.95 & 0.19 & $ 147^{+65}_{-56}$ &   & $\checkmark$ &   & \textbf{4.66} & \textbf{0.95}$\pm$\textbf{1.61} & 124.63 \\ 
 \textbf{3017360799155290368} & aIII+*!23 & 209.0703 & -19.3429 & 1.36 & 0.07 & 7.14 & 0.06 & $  -7^{+371}_{-396}$ &   & $\checkmark$ &   &   & \textbf{4.28}$\pm$\textbf{0.51} & 56.33 \\ 
 \textbf{3017367533662951936} & aII+!4 & 208.9002 & -19.3112 & 35.87 & 0.13 & 6.98 & 0.16 & $ 323^{+314}_{-123}$ &   & $\checkmark$ &   & \textbf{3.54} & \textbf{7.54}$\pm$\textbf{1.41} & \textbf{303.89} \\ 
 \textbf{3017346986540958208} & bII*!650 & 209.1877 & -19.3880 & 178.42 & 0.18 & 6.60 & 0.11 & $ 837^{+401}_{-207}$ &   &   &   & \textbf{1.53} & $10.14\pm0.98$ & 91.28 \\ 
 \textbf{3017364127743299328} & aIII+!14 & 209.0039 & -19.3738 & 1.96 & 0.01 & 6.60 & 0.01 & $  74^{+516}_{-538}$ &   & $\checkmark$ &   &   & \textbf{0.13}$\pm$\textbf{0.08} & \textbf{334.86} \\ 
\hline
\enddata

\tablenotetext{a}{from the \emph{Hipparcos Input Catalogue} \citep{Turon1992}.}
\end{deluxetable*}
\end{longrotatetable}

\cite{McBride2019} compiled a list of 5988 known Orion A complex YSO members from
the literature, with 4346 of these that are part of the \gaia DR2 catalog and
therefore part of the initial $\sim$122 million sources around $45\deg$ of the ONC.
From these, 2598 pass the clean sample criteria defined on \S\ref{sec:sample} and
432 of these were found by our 2D trace-back condition. Note that the 5988 sources
listed by \cite{McBride2019} are members of the whole Orion A complex, while the
432 sources selected by the 2D trace-back method are sources that we estimate came
from the ONC or are currently within 10\arcmin\ of it.

\begin{figure}
\centering
\includegraphics[width=\columnwidth]{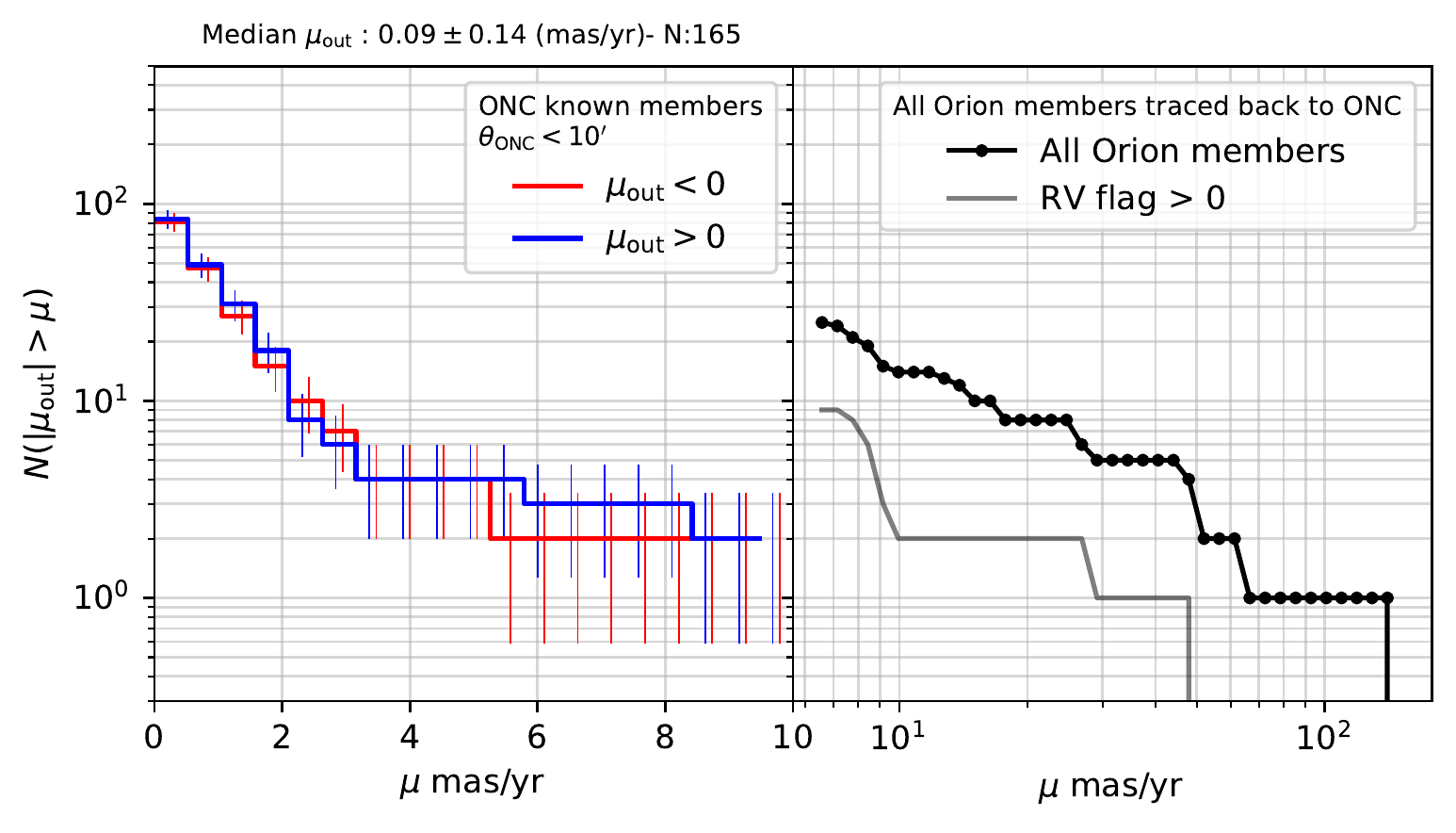}
\caption{%
        Distribution of magnitudes of radially outward ($\mu_{\rm out}>0$) and
        inward ($\mu_{\rm out}<0$) proper motions for ONC sources in the traceback
        sample within 10\arcmin, i.e., all sources within 10\arcmin.  Left panel
        shows the distribution when using only known ONC members in this region,
        from \cite{DaRio2016}. Right panel, shows the expanded high proper motion
        tail of all Orion A complex members that were traced back to the ONC (with
        $\mu_{\rm out}>0$).
}%
\label{fig:exp}
\end{figure}

We note that our 2D traceback method, in addition to finding outward moving
sources, also identifies ONC members that are currently within 1.2\,pc, i.e.,
10\arcmin, of the ONC, but moving towards the ONC center. We have measured the
radial component of the proper motion of individual stars as $\mu_{\rm out}$,
where $\mu_{\rm out}>0$ denotes the direction moving away from the ONC. After
correcting for perspective expansion caused by the radial motion of the ONC with
respect to the Sun \citep[see][]{vanLeeuwen2009,Kuhn2019}, the weighted median
radial proper motion with respect to the ONC center for sources in this region is
$0.09\pm0.14$\,\masyr when using the membership list of \cite{DaRio2016}, This
value is smaller than that measured by \cite{Kuhn2019} ($0.23\pm0.2\,\masyr$). This
may be due to the different membership list used, but also because we are only
measuring within the half mass radius of the ONC, which should be mostly populated
by bound stars. Including a wider area would also include a larger fraction of
unbound stars moving outwards. Our estimated value and the distribution shown in
the left panel of Figure~\ref{fig:exp} means there is no evidence for the expansion
(or contraction) of the ONC center. We have checked that within this region, the
ONC only shows an expansion signature when the sample is contaminated, i.e., rising
to a maximum value of $0.24\pm0.12\,\masyr$ if all sources in this region are used.

We list the properties of highest ONC-frame velocity ONC member sources in
Table~\ref{tab:known}, displaying down to 6\,\kms, which is estimated to be twice
the 2D velocity dispersion of the ONC, i.e., $\sqrt{2}\sigma_{\rm 1D}$, where
$\sigma_{\rm 1D}=2.3\,\kms$ \citep{DaRio2017}.

The first section of Table~\ref{tab:known} shows runaway candidates reported
by~\cite{McBride2019}. They found 9 sources coming from the center of the ONC, of
which we recover 7. The two sources that we do not recover, V1916 Ori and 2MASS
J05382070--0610007, were filtered out in \S\ref{sec:sample} because both have high
RUWE. We also recovered Brun 711 (aIII+!4), which was reported as a visitor of the
ONC. Given its trajectory and very high $v_{\rm r,opt}$, we agree that it likely
did not come from the ONC center. From the remaining group, 2MASS
J05351295--0417499 (bI+!4), is the fastest candidate with $v_{\rm t}= 26\,\kms$ in
the ONC frame. However, it does require a quite high radial velocity to reach the
ONC (158\,\kms).  The two sources that follow are aI+ sources, V1440 Ori (aI+2) and
CRTS J053223.9--050523 (aI+3) which pass all flags, therefore are very likely true
runaway stars. The next sources, Haro 4--379(bI+*1) and V1961 Ori (bII+9), are
scored as \emph{b} but are very close to passing the failed WYSO flag. Thus their chances
to be true YSOs are still high, and anyway more evolved YSOs, and especially
runaways, may have lost their infrared excess. Similar considerations apply to Brun
259 (bII+13), which has P(YSO) $=0.71$.  The remaining source V1321 Ori (bII+!29)
is too bright for the variability test, although it has been classified as variable
by the All Sky Automated Survey \citep{Pojmanski1998}, but has a low P(YSO)(0.21).

From the 432 ONC members selected by the 2D traceback method, 57 have a $v_{\rm
r,opt}$ higher than the 100\,\kms\ threshold. This only means that they likely did
not come from the 10\arcmin\ search area used in this work, however it does not
mean that they did not come from other regions of the ONC complex. 
The right panel of Figure~\ref{fig:exp} shows the proper motion distribution for all
Orion members that were traced back to the ONC using the 2D traceback method. The
source with the highest $\mu_{\rm out}$ is 2MASS J05430583--0807574 (bI+*!6) which
moves with $\mu_{\rm out} = 152.8\pm6.4\,\masyr$, however it requires a very high
radial velocity (-697\,\kms) to reach the ONC, and does not pass the RV flag
criterion. In fact, many Orion members do not pass the RV flag criterion, which
means that they may not have come from the ONC, but rather from other regions of the
Orion complex. We also show the $\mu_{\rm out}$ distribution for traced back Orion
members that pass the RV flag as a solid gray line of Fig.~\ref{fig:exp} right.
The two fastest sources that likely came from the ONC are actually \aeaur\ and
\mucol, that stand far from the next 3D traced back source with 9.6\,\masyr, which
is Haro 4--379 \citep[source \emph{f} in][]{McBride2019}.
\afterpage{\clearpage
\begin{longrotatetable}
    \begin{deluxetable*}{rrrrrrrrrrrrrrr}
        \tabletypesize{\scriptsize}
        \tablecaption{%
        Same as Table~\ref{tab:known}, but for sources with score \emph{a+}. The
        table is divided in 6 sections grouped by scores aI+, aII+ and aIII+ and
        the same but for sources in the ``contaminated" zone, i.e., aI+*, aII+* and
        aIII+*.  A version of this table listing all $\sim$17,000 2D traceback
        candidates will be available in the electronic version of the published
        paper.
        }
        \tablehead{
\colhead{Gaia ID}  &\colhead{Label}  &\colhead{$\ell\,(^\circ)$}  &\colhead{$b\,(^{\circ})$}  &\colhead{$d_{\rm ONC}$}  &\colhead{$\theta_{\rm ONC}$}  &\colhead{$v_{t}'$}  &\colhead{$t_{\rm back}$}  &\colhead{$v_{\rm r,opt}$}  &\colhead{$v_{\rm r}$}  &\colhead{YSO}  &\colhead{P(YSO)}  &\colhead{$V_i$}  &\colhead{$D_{\rm min}$}  &\colhead{$\theta_{\rm PA}$}  \\
\nocolhead{None} &\colhead{Score} &\colhead{[J2015.5]} &\colhead{[J2015.5]} &\colhead{(pc)} &\colhead{($^\circ$)} &\colhead{(km/s)} &\colhead{(Myr)} &\colhead{(km/s)} &\colhead{(km/s)} &\nocolhead{None} &\colhead{ } &\colhead{ } &\colhead{(')} &\colhead{($^\circ$)} \\
}
\startdata
  3209590577396377856 & aI+1 & 208.3906 & -19.6195 & 25.82 & 0.66 & 25.29 & 0.17 & $\mathbf{ -42^{+192}_{-150}}$ &   & $\checkmark$ & \textbf{0.97$^*$} & \textbf{1.69} & \textbf{1.61}$\pm$\textbf{2.04} & \textbf{249.14} \\ 
 3017304105587577728 & aI+7 & 209.3875 & -18.9411 & 6.98 & 0.60 & 4.55 & 0.54 & $\mathbf{  29^{+12}_{-11}}$ &   & $\checkmark$ & \textbf{0.85$^*$} & \textbf{2.58} & \textbf{5.84}$\pm$\textbf{4.67} & \textbf{40.55} \\ 
 3209532616814106112 & aI+17 & 208.6979 & -19.2338 & 36.67 & 0.34 & 1.79 & 0.60 & $\mathbf{  13^{+3}_{-4}}$ &   & $\checkmark$ & \textbf{0.94$^*$} & \textbf{1.43} & \textbf{9.90}$\pm$\textbf{5.10} & \textbf{295.79} \\ 
\hline
  3017366365431829248 & aII+10 & 208.9651 & -19.2519 & 1.19 & 0.14 & 2.52 & 0.19 & $\mathbf{  56^{+38}_{-28}}$ &   & $\checkmark$ &   & \textbf{2.34} & \textbf{6.47}$\pm$\textbf{1.58} & \textbf{341.78} \\ 
 3017369217290105088 & aII+17 & 208.9829 & -19.2275 & 16.65 & 0.17 & 1.89 & 0.25 & $\mathbf{  16^{+20}_{-20}}$ &   & $\checkmark$ &   & \textbf{1.61} & \textbf{2.38}$\pm$\textbf{2.25} & \textbf{350.65} \\ 
\hline
  3021115184676332288 & aIII+1 & 212.0112 & -12.5996 & 53.47 & 7.72 & 54.17 & 0.89 & $\mathbf{  28^{+6}_{-6}}$ & \textbf{31.89$\pm$0.81} & $\checkmark$ &   &   & \textbf{4.50}$\pm$\textbf{7.52} & \textbf{23.87} \\ 
 3021115180380492800 & aIII+2 & 212.0118 & -12.5984 & 62.43 & 7.72 & 50.65 & 0.89 & $\mathbf{   1^{+5}_{-5}}$ &   & $\checkmark$ &   &   & \textbf{9.63}$\pm$\textbf{7.49} & \textbf{23.87} \\ 
 3184037106827136128 & aIII+3 & 207.1251 & -27.2355 & 186.25 & 8.42 & 30.14 & 1.15 & $\mathbf{ -74^{+1}_{-1}}$ & \textbf{-26.58$\pm$16.44} & $\checkmark$ &   &   & \textbf{4.38}$\pm$\textbf{9.82} & \textbf{193.49} \\ 
 3209074803362165888 & aIII+4 & 208.6251 & -20.7815 & 11.67 & 1.51 & 11.94 & 1.08 & $\mathbf{  29^{+10}_{-10}}$ &   & $\checkmark$ &   &   & \textbf{6.73}$\pm$\textbf{9.25} & \textbf{195.35} \\ 
 3209424795953358720 & aIII+5 & 208.8879 & -19.5649 & 26.23 & 0.22 & 4.19 & 0.55 & $\mathbf{ -24^{+18}_{-18}}$ & \textbf{22.20$\pm$3.81} & $\checkmark$ &   &   & \textbf{4.33}$\pm$\textbf{4.73} & \textbf{213.73} \\ 
 3017367494996951424 & aIII+8 & 208.9083 & -19.3257 & 20.18 & 0.12 & 2.56 & 0.19 & $\mathbf{  -1^{+11}_{-14}}$ &   & $\checkmark$ &   &   & \textbf{0.99}$\pm$\textbf{1.58} & \textbf{300.16} \\ 
 3017373718415827840 & aIII+9 & 208.8754 & -19.2293 & 28.36 & 0.21 & 2.44 & 0.46 & $\mathbf{ -18^{+19}_{-38}}$ &   & $\checkmark$ &   &   & \textbf{9.69}$\pm$\textbf{3.91} & \textbf{319.27} \\ 
 3017364162103039104 & aIII+11 & 209.0026 & -19.3713 & 16.98 & 0.01 & 1.94 & 0.02 & $\mathbf{   5^{+137}_{-105}}$ &   & $\checkmark$ &   &   & \textbf{0.37}$\pm$\textbf{0.19} & \textbf{334.62} \\ 
\hline
\hline
  3015321754828860928 & aI+*1 & 212.6815 & -19.8111 & 27.73 & 3.65 & 80.31 & 0.28 & $\mathbf{  23^{+96}_{-73}}$ &   & $\checkmark$ & \textbf{0.98$^*$} & \textbf{1.74} & \textbf{2.84}$\pm$\textbf{3.04} & 96.63 \\ 
\hline
  3016780428803888768 & aII+*1 & 209.7348 & -21.1012 & 25.94 & 1.93 & 54.14 & 0.23 & $\mathbf{ -39^{+13}_{-13}}$ &   & $\checkmark$ &   & \textbf{4.10} & \textbf{1.47}$\pm$\textbf{1.95} & 157.08 \\ 
 3017265004194028416 & aII+*15 & 209.1595 & -19.4545 & 2.86 & 0.17 & 1.80 & 0.25 & $\mathbf{  46^{+52}_{-32}}$ &   & $\checkmark$ &   & \textbf{2.18} & \textbf{9.26}$\pm$\textbf{2.09} & 115.04 \\ 
\hline
  3012916087811006848 & aIII+*1 & 214.1585 & -24.1936 & 63.29 & 7.10 & 65.77 & 0.63 & $\mathbf{ -22^{+6}_{-6}}$ &   & $\checkmark$ &   &   & \textbf{6.76}$\pm$\textbf{5.29} & 133.04 \\ 
 2983790269606043648 & aIII+*2 & 218.7332 & -25.2234 & 85.11 & 11.23 & 61.02 & 1.03 & $\mathbf{   0^{+3}_{-3}}$ & \textbf{16.61$\pm$1.40} & $\checkmark$ &   &   & \textbf{5.03}$\pm$\textbf{8.67} & 120.98 \\ 
 2963542281945430400 & aIII+*3 & 227.4895 & -25.5259 & 229.65 & 18.98 & 60.27 & 0.94 & $\mathbf{ -75^{+0}_{-0}}$ & \textbf{-12.91$\pm$0.62} & $\checkmark$ &   &   & \textbf{6.14}$\pm$\textbf{7.87} & 108.38 \\ 
 3009308457018637824 & aIII+*4 & 216.0105 & -20.9758 & 66.18 & 7.08 & 54.20 & 0.72 & $\mathbf{ -19^{+6}_{-5}}$ & \textbf{-43.86$\pm$0.87} & $\checkmark$ &   &   & \textbf{7.20}$\pm$\textbf{6.07} & 102.81 \\ 
 3008883530134150016 & aIII+*5 & 216.3451 & -21.7671 & 105.38 & 7.61 & 51.24 & 0.71 & $\mathbf{ -63^{+8}_{-7}}$ & \textbf{-27.00$\pm$0.87} & $\checkmark$ &   &   & \textbf{2.73}$\pm$\textbf{6.25} & 108.00 \\ 
 2969823139038651008 & aIII+*6 & 220.6746 & -25.7842 & 132.45 & 13.11 & 51.09 & 1.18 & $\mathbf{ -33^{+1}_{-1}}$ & \textbf{6.37$\pm$1.08} & $\checkmark$ &   &   & \textbf{6.54}$\pm$\textbf{9.90} & 118.75 \\ 
 2984725369883664384 & aIII+*7 & 216.4166 & -23.9746 & 227.51 & 8.66 & 47.14 & 0.54 & $-161^{+1}_{-1}$ & \textbf{-34.64$\pm$0.14} & $\checkmark$ &   &   & \textbf{8.40}$\pm$\textbf{4.53} & 121.79 \\ 
 3017246651795678720 & aIII+*9 & 209.3164 & -19.5233 & 29.09 & 0.34 & 2.79 & 0.45 & $\mathbf{ -10^{+10}_{-11}}$ &   & $\checkmark$ &   &   & \textbf{5.62}$\pm$\textbf{3.84} & 114.35 \\ 
 3017359871442791168 & aIII+*11 & 209.0968 & -19.3799 & 24.31 & 0.09 & 2.51 & 0.11 & $\mathbf{ -66^{+49}_{-81}}$ & \textbf{25.98$\pm$7.39} & $\checkmark$ &   &   & \textbf{0.19}$\pm$\textbf{0.97} & 87.34 \\ 
\hline
\enddata
        \label{tab:bestcandidates}
    \end{deluxetable*}
\end{longrotatetable}
}

\subsection{New sources}%
\label{sec:newsources}

In this section we preset runaway candidates that fulfil various flags and that
were not flagged as members of the ONC in the literature.
Table~\ref{tab:bestcandidates} shows the best candidates from this group sectioned
by score. The first three sections show sources with score aI+, aII+ and aIII+,
sorted by their transverse velocity in the ONC frame within each section. The
following three sections show best scored sources, but that fail the position angle
flag, i.e., sources scored aI+*, aII+* and aIII+*. The trajectories of these
sources are shown in Figure~\ref{fig:bestcandidates}, color coded by their score. 

We can see in Table~\ref{tab:bestcandidates} that the best candidate in the
catalog, i.e., source aI+1, aka 2MASS J05332200--0458321, has not been flagged as
being a member of the ONC in the literature. This source moves with a transverse
velocity of 25.29\,\kms\ in the ONC frame and was ejected $\sim$170,000 years ago.
This source has been reported before by~\cite{Rebull2000} as a M6 star, 
and very recently it was also reported as runaway candidate for
\cite{Schoettler2020} who estimated an age of $5.0^{+15}_{-4}$\,Myr using
\emph{PARSEC} isochrones.  This source is at 388 parsec form the Sun and we can see
from Table~\ref{tab:bestcandidates} that its 3D distance from the ONC, $d_{\rm
ONC}$, is 25 pc.  While its transverse velocity is high, it is also quite close to
the ONC in projection, i.e., currently at $\theta_{\rm ONC} = $40'. 

The other candidates that fulfil all flags and have not previously been noted as 
members of the ONC are DR2 3017304105587577728 (aI+7), which has a transverse
velocity of 4.55\,\kms, and DR2 3209532616814106112 (aI+17), which has $v_{\rm t} =
1.79\,\kms$. Both sources have small transverse velocities within the
recent estimates of escape speed of 5.6\,\kms \citep{Kim2019}, so they are most
likely still bound to the ONC.

As discussed in \S\ref{sec:PAflag}, the PA flag shows candidates outside the zone
in the sky that is most contaminated by Galactic streaming of stars. Next, we
examine sources that are positive on all criteria, but that happen to be in this
contaminated zone, and therefore there is a higher chance that they are false positives.
There are 79 sources scored a+*, of which 32 are scored aI+*. Out of these 32
sources, only one of them is not already identified as an ONC member. This source is DR2
3015321754828860928 (aI+*1), which moves with a high transverse velocity 80.31\,\kms.
Its current position is 3.65 degrees from the ONC and it was ejected 280,000 years ago.
This source does not appear in other catalogs on Simbad, but given its luminosity is
very likely to be a low-mass star. {This source was also selected by the
recent work of \citep{Schoettler2020}, who estimated an age of
$0.4^{1.1}_{-0.3}$\,Myr.}

\subsection{Sources with measured radial velocities}%
\label{sec:rvcandidates}

About 7\% of sources in our sample have measured radial velocities. None of them
have been scored with the maximum score of aI+. The best source found with
available radial velocity is source aII4 (2MASS J05343170--0351513), which is a
known ONC YSO \citep{Megeath2012}, with a radial velocity of 20.63\,\kms, which is
within the range of the estimated $v_{\rm r,opt}$ distribution to reach the ONC
center, i.e., the range 19 -- 27\,\kms.  We note that this source is very close to
passing the CA flag with $\dmin=7.26\pm 10.37\arcmin$ and therefore it likely came
from the ONC.

There are three previously unknown members of the ONC scored aIII+ and 7 in the
contaminated zone with score aIII+*. Table \ref{tab:rv} summarizes the
properties of a+ candidates with available radial velocities.  All these sources
are too bright to be evaluated by variability and were not available on \allwise.
Source aIII+3 (TYC 4762--492--1), is the source with one of longest traceback time
of the group with $t_{\rm back}=1.15$\,Myr.  However, when tracing back in 3D
space, we obtained a traceback time $\tau_{\rm min,3D} = 2.6\pm0.2$\,Myr. Its
closest approach to the ONC in 3D is $42.5\pm9.4$\,pc and is currently at 186.2\,pc
from the ONC. Other sources with long $\tau_{\rm min,3D}$ are sources aIII+*7 (HD
36343, $2.4\pm0.1$\,Myr) and aIII+*3 (CD-23 2974, $2.4\pm0.03$\,Myr). Both of these
are already classified as high proper motion stars, but with the radial velocity
measured by \gaia, we estimate their trajectories approach to the ONC as close as
$91\pm6.6$\,pc and $97\pm4.7$~pc respectively when using straight lines
trajectories. A more comprehensive analysis, including allowance for the effects of
nonuniform motion, would be needed to discard or confirm such sources. 

Two sources in this list are also identified by \cite{Schoettler2020} as visitors
of the ONC, given their isochronal age estimates ($>20$\,Myr). These are sources
aIII+1 and aIII+*4 (TYC 5354-1317-1). We confirm that the trajectories of these
sources overlap with that of the ONC, and we estimate their closest approach
distances are $0.8\pm6.5$\,pc and $15.8\pm10.4$\,pc, respectively. Source aIII+1
does not appear in other catalogs on Simbad. 

\subsection{Color-magnitude diagram}
\label{sec:hrdiag}
 
\begin{figure*}
    \centering
    \includegraphics[width=\textwidth]{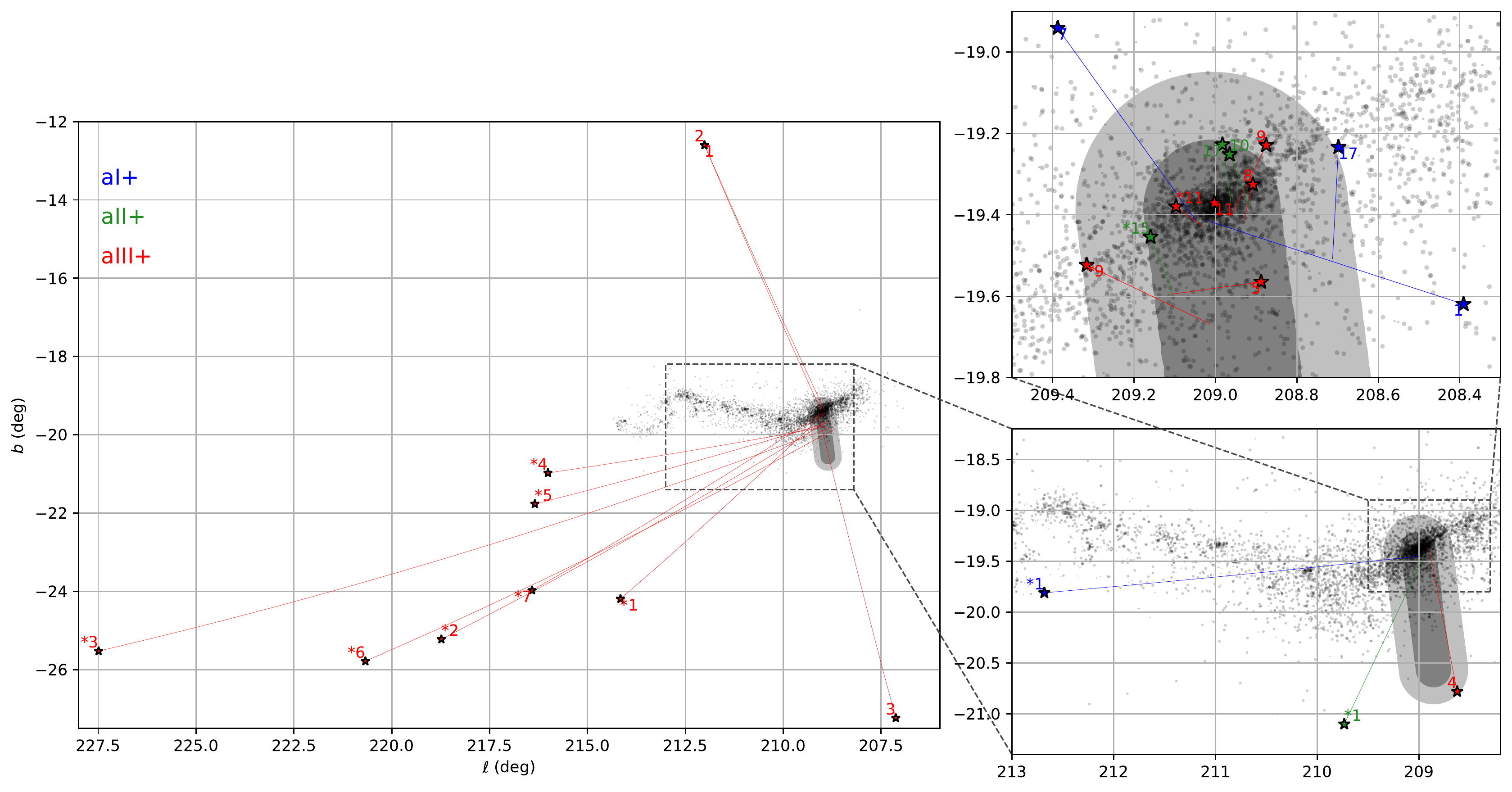}
    \caption{%
            Position and 2D trace back trajectories of candidates shown in
            Table~\ref{tab:bestcandidates}.  Gray shaded area shows the 10\arcmin\
            search threshold around the ONC and its estimated trajectory for the
            past 3 Myr.  With a larger shaded area showing the 20\arcmin\ region
            for reference. Symbol colors shows the different scores with AI+
            (blue), aII+ (green) and aII+(red).  ONC members from the literature
            \citep{McBride2019} are shown in black as reference.  For clarity, each
            source is drawn once, i.e., is only shown in one of the panels.
    }%
    \label{fig:bestcandidates}
\end{figure*}

\begin{figure*}
    \centering
    \includegraphics[width=\textwidth]{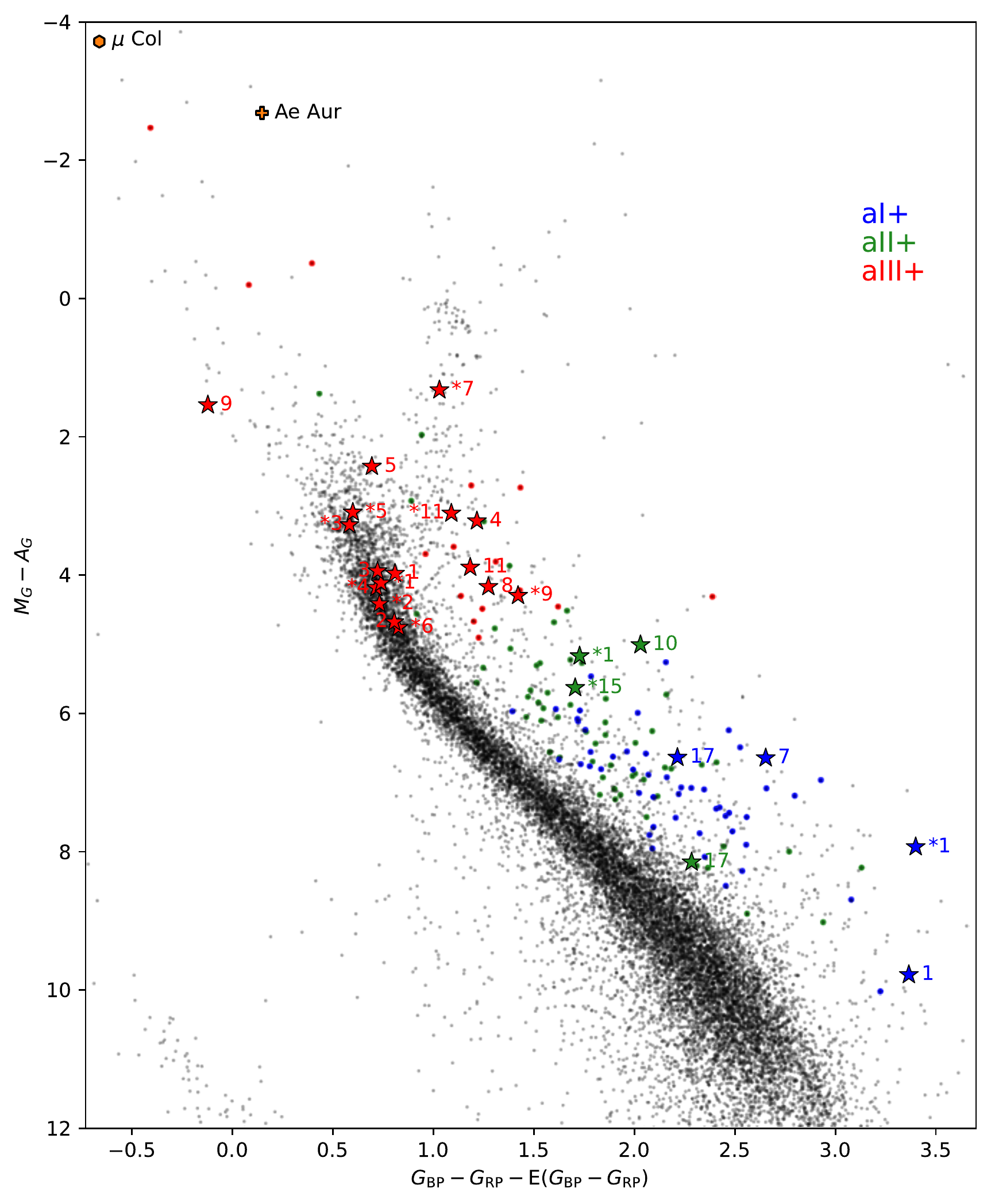}
    \caption{Color magnitude diagram for sources selected by the 2D traceback
    method (black). All sources with scores \emph{aI+}, \emph{aII+} and \emph{aIII+} 
    are highlighted in blue, green and red respectively. Star symbols shows sources
    not previously marked as Orion members, from Table~\ref{tab:bestcandidates}.}
    \label{fig:hrdiag}
\end{figure*}

Valuable insight can be obtained by examining the position in the HR diagram of the
best scored candidates discussed in the previous sections. Figure~\ref{fig:hrdiag}
shows the color-magnitude diagram using the \gaia\ photometry system, corrected by
extinction and color excess as described in \S\ref{sec:ysoflag}. Black circles show
the $\sim$17,000 sources selected by the 2D traceback method. We have highlighted
the best scored candidates grouped in aI+, aII+ and aIII+, including sources that
fail the PA flag and known members of the Orion A complex. We see that in general
aI+ and aII+ new candidates (star symbols with labels) are consistent with the rest
of ONC members with the same score, while the aIII+ group contains a group of new
candidates that populate the bright end of the main sequence, which is not
populated by Orion members in the same group.  Most of this group of sources are
currently at least several degrees away from the ONC (see
Figure~\ref{fig:bestcandidates}).

The brightest new source is HD 36343 (aIII+*7), a high proper motion star with
available radial velocity not yet associated to any parent cluster. We estimated a
closest approach to the ONC, using linear 3D traceback, of $91\pm6.6$\,pc about 2.4
Myr ago. A more comprehensive method will be needed to confirm or discard this
source.  The next brightest new source is  V* V566 Ori (aIII+9), a variable star
not marked as member of the Orion A complex, but that is just 12\arcmin\ away from
the ONC, but 28.4 parsecs away in terms of 3D distance. While it moves at low speed
(2.44\,\kms) it could still have originated from the ONC given its distance and
missing radial velocity.  Source Brun 252 (aIII+5) is quite close to the main
sequence, but has been classified as an irregular variable
\citep{Rodriguez-Ledesma2009}.

\section{High velocity distribution of the ONC}\label{sec:theory}

\subsection{Method of estimating the observed distribution}

\begin{figure}
    \centering
    \includegraphics[width=\columnwidth]{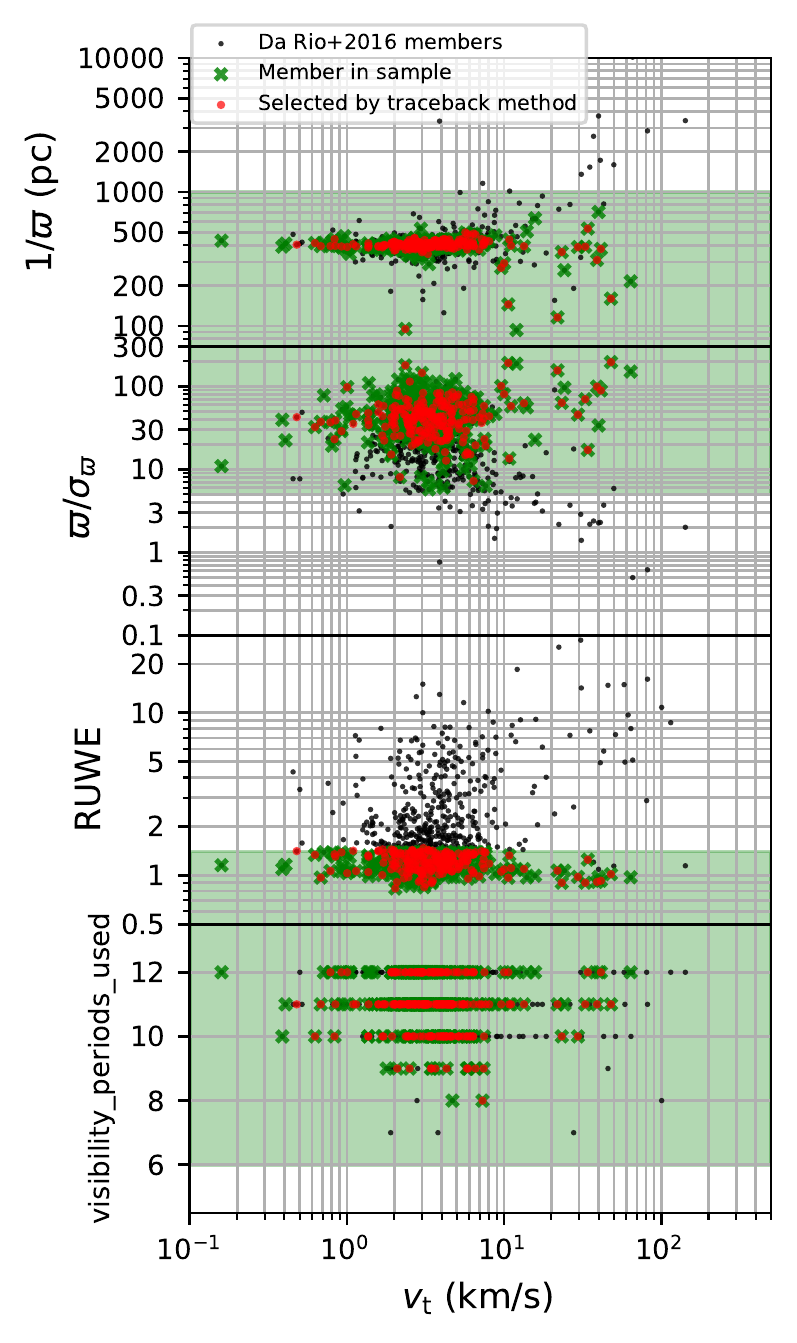}
    \caption{%
     Transverse velocity of ONC members within 20\arcmin\ under the different quality
     thresholds used in this work. Green areas shows the selection criteria for
     each quantity.  Sources in green are part of the clean sample where the
     traceback method was applied.  Red sources, are the ones selected by the
     traceback method.}
\label{fig:cleanmembers}
\end{figure}

\cite{Farias2019} discussed how the statistics of the unbound stellar population
created by dynamical ejection events can be an important constraint on star cluster
formation models \citep[see also][]{Schoettler2019}. However, to obtain such
constraints one needs to be careful about how models are compared to observational
data. While masses and ages of individual stars are challenging to determine,
individual velocities rely only on astrometric measurements, such as proper motion
and parallax. Unfortunately we do not have radial velocities for most of the
sources from \gaia. However, with the unprecedented accuracy of parallax and proper
motion measurements, we can obtain reliable transverse velocities for the ONC
members and candidate runaways.  In this section, we construct the high-velocity
distribution of the ONC. We start this analysis, by taking the membership list used
in \cite{DaRio2016}. Since such a membership list contains sources from the whole
Orion complex, we select sources within 20\arcmin\ around the core of the ONC,
which counts 740 sources flagged as members by \cite{DaRio2016} and that are also
observed by \gaia. Figure~\ref{fig:cleanmembers} shows the transverse velocity for
these sources versus the different quality criteria described in
\S\ref{sec:sample}. Most of these sources do not pass all quality criteria (shown
as green areas), but the 336 sources that pass (green symbols) are part of the
initial sample. From these 336 ONC members in the clean sample, 220 are captured by
the 2D traceback method described in \S\ref{sec:sample}. 

Since not all of these sources have reliable astrometry, we used the distance
estimations of \cite{Bailer-Jones2018} in order to calculate transverse velocities.
The resulting velocities do not change significantly for the sources that pass the
quality cuts, but this step improves results considerably for sources with
relatively low quality.  As can be seen in Figure~\ref{fig:cleanmembers}, most of
the sources with $v_{\rm t}>10\,\kms$ have large errors in their parallax
($\sigma_{\varpi}/\varpi > 0.2$), as well as large RUWE, which means that the
observations are not consistent with the astrometric model used by \gaia.
Therefore, most of the sources with high $v_{\rm t}$ may be caused by uncertainties
in astrometry and we exclude such sources from the sample.

Figure~\ref{fig:vdist} shows the distribution of sources with high $v_{\rm t} >
v_{\rm min}$.  The distribution obtained when using only known ONC members that
were traced-back using the method described in \S\ref{sec:sample} is shown in red.
There are 17 sources with velocities above 10\,\kms\ in this sample with a rapid
decrease in numbers as the velocity cut increases, with the fastest source V* V1175
Ori $v_{\rm t} = 47\,\kms$.

\begin{deluxetable*}{rrrrrrrrrrr}
        \caption{%
        Sources that fullfil the RV flag with available \gaia\ DR2
        radial velocities with scores \emph{A+}. \gaia\ ID in bold indicate sources
        that are known members of the Orion A complex. $v_{\rm 3D,ONC}$ shows the
        space velocity in the frame of reference of the ONC. $\tau_{\rm min,3D}$ is
        the traceback time obtained by 3D traceback and $D_{\rm min,3d}$ the
        corresponding 3D closest approach to the ONC.%
        \label{tab:rv}}
\tablehead{
\colhead{Gaia ID}  &\colhead{Label}  &\colhead{$\ell\,(^\circ)$}  &\colhead{$b\,(^{\circ})$}  &\colhead{distance}  &\colhead{$d_{\rm ONC}$}  &\colhead{$\theta_{\rm ONC}$}  &\colhead{$v_{\rm 3D,ONC}$}  &\colhead{$v_{\rm r}$}  &\colhead{$\tau_{\rm min,3d}$}  &\colhead{$D_{\rm min,3d}$}  \\
\nocolhead{None} &\colhead{Score} &\colhead{[J2015.5]} &\colhead{[J2015.5]} &\colhead{(pc)} &\colhead{(pc)} &\colhead{($^\circ$)} &\colhead{(Myr)} &\colhead{(km/s)} &\colhead{(Myr)} &\colhead{(pc)} \\
}
\startdata
  2983790269606043648 & aIII+*2 & 218.7332 & -25.2234 & $ 370^{+4}_{-4}$ & 85.11 & 11.23 & 67.83 & \textbf{16.61$\pm$1.40} & $ 1.1\pm 0.1$ & \textbf{14.4}$\pm$\textbf{ 7.9} \\ 
 2963542281945430400 & aIII+*3 & 227.4895 & -25.5259 & $ 203^{+1}_{-1}$ & 229.65 & 18.98 & 79.76 & \textbf{-12.91$\pm$0.62} & $ 2.4\pm 0.0$ & \textbf{97.2}$\pm$\textbf{ 4.7} \\ 
 3009308457018637824 & aIII+*4 & 216.0105 & -20.9758 & $ 366^{+5}_{-4}$ & 66.18 & 7.08 & 91.72 & \textbf{-43.86$\pm$0.87} & $ 0.6\pm 0.1$ & \textbf{15.0}$\pm$\textbf{10.4} \\ 
 3021115184676332288 & aIII+1 & 212.0112 & -12.5996 & $ 404^{+5}_{-5}$ & 53.47 & 7.72 & 57.83 & \textbf{31.89$\pm$0.81} & $ 0.9\pm 0.1$ & \textbf{ 0.8}$\pm$\textbf{ 6.5} \\ 
 3008883530134150016 & aIII+*5 & 216.3451 & -21.7671 & $ 319^{+9}_{-9}$ & 105.38 & 7.61 & 77.44 & \textbf{-27.00$\pm$0.87} & $ 1.2\pm 0.2$ & \textbf{22.5}$\pm$\textbf{19.4} \\ 
 2969823139038651008 & aIII+*6 & 220.6746 & -25.7842 & $ 306^{+2}_{-2}$ & 132.45 & 13.11 & 61.34 & \textbf{6.37$\pm$1.08} & $ 1.8\pm 0.0$ & \textbf{53.9}$\pm$\textbf{ 7.0} \\ 
 2984725369883664384 & aIII+*7 & 216.4166 & -23.9746 & $ 190^{+1}_{-1}$ & 227.51 & 8.66 & 80.38 & \textbf{-34.64$\pm$0.14} & $ 2.4\pm 0.1$ & \textbf{91.1}$\pm$\textbf{ 6.6} \\ 
 3184037106827136128 & aIII+3 & 207.1251 & -27.2355 & $ 232^{+2}_{-2}$ & 186.25 & 8.42 & 63.01 & \textbf{-26.58$\pm$16.44} & $ 2.6\pm 0.2$ & \textbf{42.5}$\pm$\textbf{ 9.4} \\ 
 3209424795953358720 & aIII+5 & 208.8879 & -19.5649 & $ 387^{+6}_{-6}$ & 26.23 & 0.22 & 6.18 & \textbf{22.20$\pm$3.81} & $ 2.0\pm 1.7$ & \textbf{ 8.2}$\pm$\textbf{ 8.1} \\ 
 \textbf{3017364028971010432} & aIII+6 & 209.0086 & -19.4010 & $ 401^{+8}_{-7}$ & 12.32 & 0.02 & 12.04 & \textbf{15.77$\pm$12.22} & $ 0.1\pm 0.9$ & \textbf{ 0.5}$\pm$\textbf{11.7} \\ 
 \textbf{3017367391918532992} & aIII+7 & 208.9169 & -19.2701 & $ 376^{+5}_{-4}$ & 37.61 & 0.15 & 10.02 & \textbf{17.28$\pm$9.08} & $ 2.6\pm 1.0$ & \textbf{ 4.7}$\pm$\textbf{11.2} \\ 
 3017359871442791168 & aIII+*11 & 209.0968 & -19.3799 & $ 389^{+8}_{-8}$ & 24.31 & 0.09 & 3.54 & \textbf{25.98$\pm$7.39} & $ 1.3\pm 3.0$ & \textbf{12.5}$\pm$\textbf{12.6} \\ 
 \textbf{3017358978089804672} & aIII+*12 & 209.1422 & -19.4126 & $ 396^{+8}_{-8}$ & 17.28 & 0.14 & 5.01 & \textbf{23.14$\pm$14.82} & $ 1.1\pm 4.2$ & \textbf{ 3.2}$\pm$\textbf{13.6} \\ 
 \textbf{3017364544367271936} & aIII+10 & 208.9696 & -19.4842 & $ 395^{+4}_{-4}$ & 18.35 & 0.11 & 5.11 & \textbf{22.38$\pm$10.35} & $ 1.2\pm 1.6$ & \textbf{ 3.6}$\pm$\textbf{ 6.2} \\ 
 \textbf{3209521037582290304} & aIII+13 & 208.8633 & -19.3789 & $ 392^{+8}_{-8}$ & 21.52 & 0.14 & 22.97 & \textbf{4.16$\pm$9.09} & $ 0.4\pm 0.6$ & \textbf{ 1.2}$\pm$\textbf{11.8} \\ 
 \textbf{3017360554330360320} & aIII+*13 & 209.0450 & -19.4281 & $ 388^{+7}_{-7}$ & 25.61 & 0.06 & 7.03 & \textbf{20.33$\pm$13.98} & $ 2.0\pm 3.6$ & \textbf{ 3.9}$\pm$\textbf{20.7} \\ 
\hline
\enddata

\end{deluxetable*}

We have also included the corresponding distributions using the best-scored
candidate runaways from this work. However, before describing such distributions,
we first discuss some possible systematic uncertainties due to contaminants in our
sample.  As mentioned in previous sections, there is a region in the sky, which we
have characterized as being from PA = 50$^\circ$ to 160$^\circ$, i.e., pointing
approximately parallel to the Galactic plane, where most projected Galactic orbits
appear to move away from the ONC, causing an asymmetry on traceback selection and
increased levels of contamination of the sample with false positives. In order to
see how this contamination affects the constructed velocity distributions, we
proceed with the following analysis using two approaches: Method 1: construct the
velocity distribution ignoring the effects of this higher contamination zone;
Method 2: construct the velocity distribution by using only the region of the sky
outside this contamination zone, i.e., using only sources that pass the PA flag.
Then final numbers are boosted by a statistical correction factor of 1.44 that
accounts for the missing region of the sky that was not considered.  During the
following description, the left column of panels in Figure~\ref{fig:vdist} shows
results using Method 1, while the right column shows results using Method 2.

Starting with Method 1, in order to include the best candidates in the sample, we
first examine all \emph{a} scored candidates that pass the RV flag, i.e., with
scores aI(*), aII(*), aIII(*), which includes sources that fail the PA flag. We can
see from this distribution that even this sample is still likely to be highly
contaminated with false positives (see top left panel in Figure~\ref{fig:vdist}).
We consider that this distribution is unreliable for representing the true ONC high
velocity population.

We next constructed another velocity distribution using only the best candidates
from our sample, i.e., the sources with scores \emph{aI}, but still also including sources
that fail the PA flag that pass the RV criteria. This method selects a smaller
fraction of candidates (black filled circles on Figure~\ref{fig:vdist}), which are
more likely to be true runaways.

Then, the Method 1 estimate for the sample for the high velocity tail is composed
by a combination of our best candidates (95 sources),  272 traced back members from
\cite{DaRio2016} that pass all the astrometric quality criteria (red filled circles
on Figure~\ref{fig:cleanmembers}) and we have also added \aeaur\ and \mucol\ to the
final sample, that, given some overlap, has 314 sources.  We show this profile as a
solid line in the top left panel of Figure~\ref{fig:vdist}, which is the
combination of the open red and black filled circles with the addition of \mucol\
and \aeaur\ that are not part of either sample.

The Method 2 estimate follows the same procedure as Method 1, described above,
except excluding sources from the Galactic streaming contaminated zone.  With this
removal of sources that fail the PA flag, the total number of sources drops to 126.
Once boosted by the correction factor of 1.44, the estimated number of sources in
the high velocity distribution is 181. 

\subsection{Comparison with theoretical models}\label{sec:models}

\begin{figure*}
    \centering
    \includegraphics[width=0.76\textwidth]{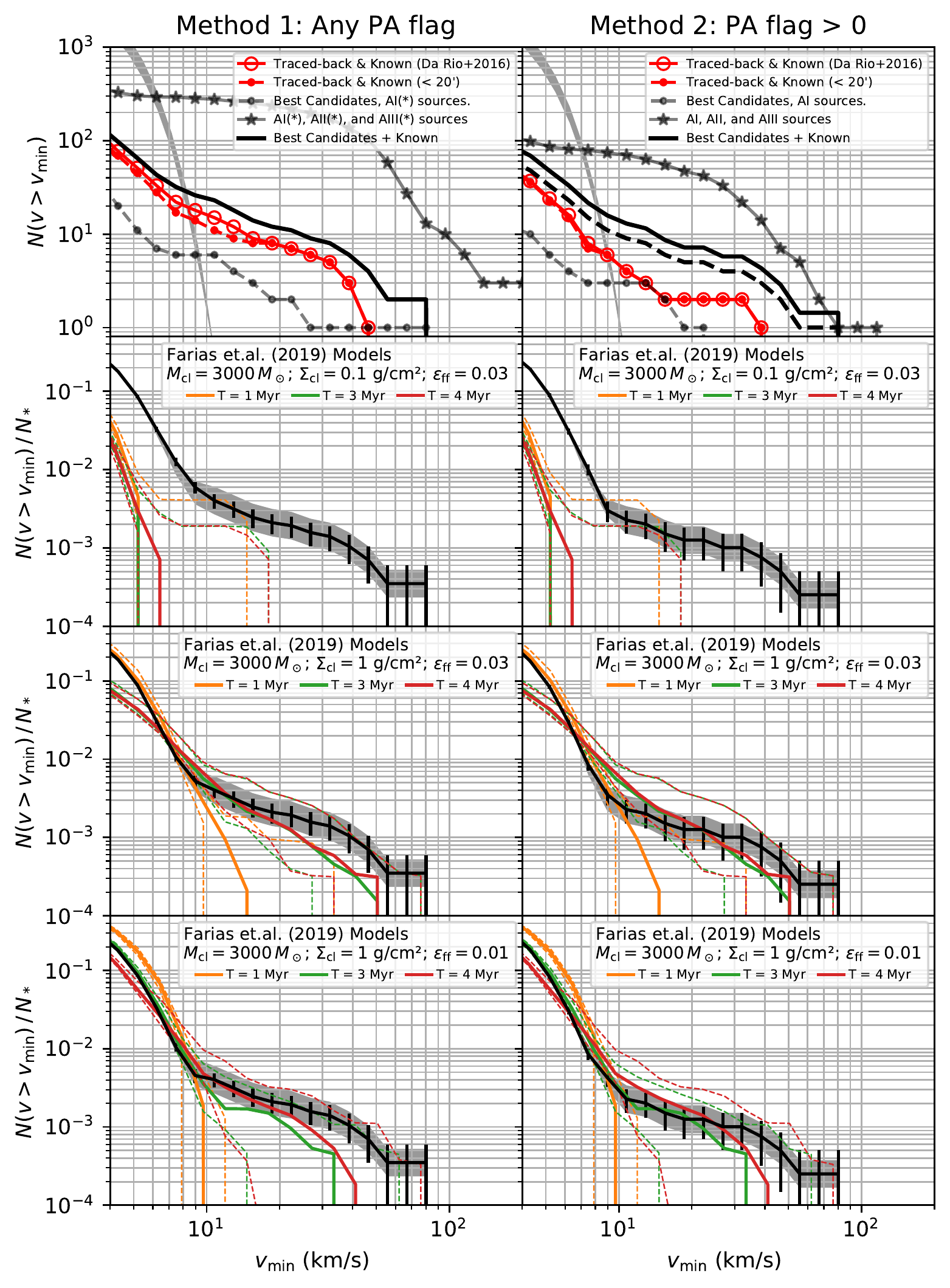}
    \caption{ Transverse (plane-of-sky) velocity distributions for ONC members, new
            candidates from this work and numerical models. We compare the
            constructed high velocity distribution when ignoring the Galactic
            streaming contamination zone (left column) and when only considering
            sources outside this zone, i.e., PA flag $>$ 0 (right column).  Top
            panel shows the number of sources with velocities above $v_{\rm min}$
            for sources that were selected using the traceback method described in
            \S\ref{sec:sample}. Black circles show aI scored candidates, where the
            (*) character indicates when including PA flag $<$ 0 sources. Star
            symbols show all sources scored \emph{a} that pass the RV flag.  Red
            open circles show all selected sources that are flagged as members in
            \cite{DaRio2016}, while filled circles show the sub-sample within
            20\arcmin\ of the ONC.  The solid black line shows the combination of
            the traceback + \cite{DaRio2016}-matched sample (red open circles) and
            new traceback candidates scored aI.  In the top right panel, this black
            line includes a correction factor of 1.44 to account for the sky area
            not considered by Method 2 (dashed line shows the distribution before
            this correction).  These solid lines in the top row are used in the
            bottom panels to compare with models.  Top second to bottom rows show
            the same metrics as in the top row, but now normalized by the number of
            members of the sample, $\Nsys$.  The black line with Poisson errors
            shows the case assuming $\Nsys\approx5700$, with grey shaded area
            resulting from the range $\Nsys\approx$ 2800 to 8600 (see text). In
            order to reach such normalizations, observations were complemented with
            velocities drawn from a Maxwell-Boltzmann velocity distribution with a
            $\sigma_{\rm 1D}=$ 2.3\,\kms\ \cite{DaRio2016} (this is shown in the
            top panel as a shaded area).  Colored solid lines show star cluster
            simulations from \citep{Farias2019} with an unresolved binary
            population (see text) at different evolutionary times.  Values are
            medians over 20 simulation realizations with corresponding 16th and
            84th percentiles as dashed lines.  }
    \label{fig:vdist}
\end{figure*}

\cite{Farias2019} conducted a series of numerical simulations of star cluster
formation, which have some properties that are similar to those expected of the
ONC. Idealised molecular clumps bounded by environments with different mass surface
densities were evolved from the earliest, gas-dominated stages, via gradual
formation of stars at various rates, until the systems were completely gas-free.
Modeling was conducted until 20\,Myr after the exhaustion of their natal gas. We
have considered various examples of these models and compared their velocity
distributions to that derived for the ONC.

The selected set of models are 3000\,$\MSun$ molecular clumps, approximated as
singular polytropic spheres \citep{McKee2003}, with a constant overall star
formation efficiency of $\epsilon=0.5$ and different star formation efficiencies
per global average free-fall time ($\epsilon_{\rm ff}$). Models include 50\%
primordial binaries. Stellar evolution, including supernovae velocity kicks, is
included. Simulations were performed using \texttt{Nbody6++}
\citep{Aarseth2003,Wang2015}, but modified to allow the gradual assembly of star
clusters including primordial binaries and a custom background potential to emulate
the influence of the background gas.  Depending on the model parameters, i.e.,
molecular cloud mass ($M_{\rm cl}$), overall star formation efficiency, surrounding
cloud mass surface density ($\Sigma_{\rm cl}$) and $\epsilon_{\rm ff}$, star
cluster formation spans over a wide range of timescales. We have selected three
sets of models that are are finished with their star formation at 1, 3 and 6 Myr.
First, a low density model with $\Sigma_{\rm cl}=0.1$\,\gcs, and $\epsilon_{\rm
ff}=0.03$ in which the cluster forms over 6.5\,Myr. Second, a model in which the
surface density is increased 10 times, $\Sigma_{\rm cl}=1.0\,\gcs$, but with the
same $\epsilon_{\rm ff}=0.03$. In this case the star formation takes 1.2\,Myr. For
the third model, the mass surface density is also $\Sigma_{\rm cl}=1.0\,\gcs$, but
$\epsilon_{\rm ff}$ is smaller, i.e., $\epsilon_{\rm ff}=0.01$, so that star
cluster formation takes place over 3.35\,Myr \cite[see][for further
details]{Farias2019}. 

Using these models we have constructed a 2D velocity distribution in order to
compare with the ONC. We have constructed this 2D velocity distribution by
discarding one of the velocity components in the simulations using snapshots at 1,
3 and 4\,Myr. In general, we expect the high velocity population to grow as time
evolves, i.e., after there has been more time for close stellar encounters leading
to dynamical ejections. The bottom panel of Figure~\ref{fig:vdist} shows the
obtained distributions for these models.  Distributions are medians of 20
realizations for each model, normalized by the current total number of member
stars. The 16th and 84th percentiles are shown as the corresponding dashed lines.
Note that these simulations have 50\% primordial binaries and the binary fraction
does not change significantly during the evolution of the cluster. For our
comparison, each binary pair is counted as one single star, as it would be if such
systems are not resolved, and velocities shown in the distribution are obtained
from the center of mass velocities from each pair. 

We compare these results with the obtained observational distributions in the
previous section (solid black lines in the top panels). The total number of members
of the ONC is uncertain. However for this comparison we use the total stellar mass
within 3\,pc estimated by \cite{DaRio2014}, of 3000\,\MSun. Using a canonical mass
function \citep{KroupaIMF} within a range between 0.01 to 100\,\MSun, the mean
stellar mass is $\langle m_i \rangle\approx0.35\,\MSun$. However, if we assume a
binary fraction $f_{\rm bin} = 0.5$ and ignore higher order multiples, the mean
mass per system is \citep[see][]{Farias2017} $\langle m_{\rm s} \rangle = \langle
m_i\rangle (1+f_{\rm bin})=0.525\,\MSun$. Then the number of stars including
unresolved binaries in the ONC with an estimated stellar mass of 3000\,\MSun is
$\Nsys=5714$. We normalize the observed distribution by the $\Nsys$ obtained
assuming a total stellar mass between 2000 to 4500\,\MSun, i.e., with
$2857<\Nsys<8571$. In order to reach such numbers from the samples of 316 and 181
members and candidates, we complement such samples with transverse velocities drawn
from a Maxwell-Boltzmann distribution with $\sigma_{\rm 1D}=2.3\,\kms$ as it has
been estimated in the ONC by various authors \citep{Jones1988,DaRio2016,Kim2019}.
We show the complementary distributions used in the top panels of
Figure~\ref{fig:vdist}, which yields the grey shaded area in the lower panels of
Figure~\ref{fig:vdist}. Note that since our sample has been initially
cleaned using the RUWE parameter, it could potentially be missing a fraction of
unresolved binaries that are ONC members.  In the simulations we have included
unresolved binaries in the distributions. However we have checked that in the
ejected population above 10\,\kms the fraction of binaries is below 5\%, and
therefore we expect that any potential discrepancy due to this method has a small
effect.

The second row panels in Figure~\ref{fig:vdist} show the comparison with the low
density cluster models ($\Sigma_{\rm cl}=0.1\,\gcs$), where we can see that they
are not able to produce enough high velocity runaways during such early phases.
Also, the velocity dispersion of these simulated clusters is much smaller than that
of the ONC. 

In the third row of panels we see that these denser clusters are able to produce a
similar number of high velocity stars, i.e., above 10\,\kms, as inferred in the
ONC. However, given the shorter dynamical times of these models, star formation is
exhausted at 1.2\,Myr. During this phase, when background gas is still present, the
velocity dispersions of the simulated clusters are quite constant. However, as soon
as the gas is depleted, the star cluster expands and the velocity dispersion drops,
as can be seen in the low velocity end of the distribution. However, at these later
times most of the strongest dynamical processing has already taken place and so the
high velocity tail does not change too much after this.  In the final set of models
shown in the bottom row panels, star formation is less efficient and star cluster
formation takes longer. While these models have a similar velocity distribution as
the previous case, they are still forming stars at 3\,Myr. In this respect they are
a better fit to the proposed age of the ONC. However, we note that these clusters
have half-mass radii of $\sim0.3$~pc at 3 to 4~Myr, which is a few times smaller
than that of the ONC.

In this third set of simulations at 3\,Myr the median velocity distribution is
quite similar to the one estimated for the ONC, especially via Method 2, although
the simulated distributions on average tend to be less populated at the highest
velocities above 45~\kms. Still, even here the discrepancy is minor, considering
the distribution in the simulations (i.e., the line of the 84th percentile) and the
Poisson sampling uncertainties in the observed distribution.

Considering the global average number densities of systems inside the half-mass
radius, these last set of high density simulations have a few 1000 stars\,pc$^3$
during the first $\sim0.5$\,Myr, rising to $\sim10^4$ by 2.5\,Myr and then
declining significantly during the next few Myr, i.e., once the gas is exhausted
and the cluster expands \citep[see Figure 7 at][]{Farias2019}. Such properties may
be quite similar to those expected during the formation and evolution of the ONC.
However, it should be emphasized that these simulations have not been designed to
match properties of the ONC. We expect within the possible parameter space of the
models there will be clusters that can be a better match. The models are also
relatively simple in that they so far do not assume any global elongation or
spatial or kinematic subclustering of stars when they are born. Other assumptions
to be investigated include effects of different initial dynamical states (currently
the initial clump is in approximate virial equilibrium, including near
equipartition magnetic fields), different degrees of primordial mass segregation
(currently none is assumed) and the effects of primordial triples and higher order
multiples (currently there are none).  Overall, these observational measurements
and comparisons provide a baseline to help develop and study the influence of these
new ingredients in the star cluster models. 

\section{Discussion \& Conclusions}

Using the unprecedented astrometric accuracy of \gaia\ DR2, we examined a wide
region extending out to 45\deg\ around the ONC to search for runaway (or walkaway)
stars it may have produced.  Within this area we have selected \Ntraceback\ sources
that have a 2D trajectory in the sky that brings them back to the ONC in the recent
past. Most of these are expected to be contaminants. Thus, using different criteria
based on signatures of youth and astrometric accuracy, we have developed a scoring
system that allowed us to filter and select the most likely runaway candidates.  In
particular, we have selected a set of 25 candidates that do not fail any of the
tests, except for being in the zone of most contamination from Galactic streaming. 
These have not been associated with the Orion A complex or the ONC previously. Six
of these sources pass all the signatures of youth tested in this work, i.e.,
variability, color-magnitude selection and IR YSO classification, and the fastest
is escaping with a transverse velocity of 65\,\kms\ in the frame of reference of
the ONC.  From the $\sim$1200 sources in the sample with available radial
velocities 491 pass the most stringent radial velocity criterion to achieve 3D
traceback, from which a small sample of 10 new sources (a subset of the 25 found
above) do not fail any signature of youth flag, and therefore are strong runaway
candidates.

Within the traceback sources, we have examined already known members of the Orion A
complex. Since the traceback method selected sources whose trajectories were
consistent with the one of the ONC within its half mass radius as a threshold,
current sources within this limit were all selected. We have examined the outward
proper motions within this area and found no signatures of radial expansion of the
cluster center, with an outward median proper motion of $0.09\pm0.014\,\masyr$
($0.18\pm0.03\,\kms$), when using only literature members in the measurement. While
membership classification is somewhat challenging and a fraction of these sources
may be non-member contaminants, we have shown that even if we use the whole sample,
with reliable astrometric measurements, outward motions can only rise to a value of
$0.23\pm0.12\,\masyr (0.44\pm0.24\,\kms)$, which is still not a clear sign of
cluster expansion. This result suggests that the ONC center may have already
reached dynamical equilibrium supporting the idea of a dynamically old system, even
though it is still in the process of forming stars \citep[see
also][]{DaRio2016,DaRio2017}.

Since the vast majority of sources in our traceback sample do not have measured
radial velocities, we have computed the distributions of radial velocities that
minimize the closest 3D trajectory to the ONC. We have obtained these distributions
by sampling each astrometric quantity within their errors using a Monte Carlo
approach. This derived sub-product of this work can be very useful for quickly
distinguishing good candidate runaways when new RV measurements are obtained in the
future. However, even though we have only used sources with the best astrometry,
there are still a considerable group of sources whose range of optimal velocities
are extremely large, and for which this metric does not provide very strong
constraints on an ejection scenario. The selection criterion we used with this
quantity was then rather restrictive, since we marked as negative sources with
required radial velocity above 100\,\kms. Such a restrictive threshold was designed
to eliminate most false positives, however some runaway stars with larger
velocities are still possible and may still be hidden in our sample.

We have also estimated the total high velocity distribution of the ONC using the
known and new members with the best astrometry and membership probability,
selecting a sample of about 200 to 300 sources, depending on the method.  While
there are still significant systematic uncertainties in the estimation of this
distribution, we have compared it with theoretical models based on simulations that
include realistic fractions of primordial binaries and gradual formation of stars,
which are necessary components for accurately capturing rates of dynamical
ejections. These simulations can successfully reproduce the normalization and shape
of the estimated velocity distribution of the ONC, however only when using higher
density models ($\Sigma_{\rm cl}=1\gcs$) with relatively slow star formation
($\epsilon_{\rm ff}=0.01$).  A more general exploration of simulation parameter
space for clusters specifically designed to match the ONC is the next step to be
conducted.

Very recently we have become aware of a parallel work from \citep{Schoettler2020},
who have performed a similar observational search for runaway stars from the ONC,
but restricting to a smaller region extending only 100~pc away, i.e., up to 14\deg
from the cluster.  They also performed N-body simulations of clusters, which in
their case were initialised with sub-virial, fractal distributions of stars with a
primordial binary fraction that depends on primary mass ($\sim$50\% on average).
While the ultimate goal of their study is similar to this work, their methods of
selection and classification of sources are very different. Some of the main
differences are that they base the classification in the comparison with the
traceback time and the estimated ages for their sources using isochrone fitting
from \emph{PARSEC} \citep{Bressan2012}. They were quite restrictive on the age of
the ONC, excluding from the runaway lists sources with ages larger than 4\,Myr.
While this is a reasonable limit for most ONC stars \citep{DaRio2016}, we consider
that it may be too restrictive a choice given the uncertainties in age estimates of
individual stars from isochrone fitting.  \citet{Schoettler2020} have reported 54
walkaway candidates with velocities between 10--30\,\kms\ and 31 runaway candidates
with velocities above 30\,\kms.  From their sources, we recovered 22 out of 31
runaways and 32 out of their 54 walkaways, but give them varying degrees of
likelihood of being true runaways.  On the other side, from our 25 best candidates
they have listed 4 as being ejected from the ONC and 2 as visitors.
These differences in the number of selected sources, given that we observed a wider
area, likely arise from the fact that we do a more restrictive initial astrometric
cleaning and that we also use a more restrictive baseline boundary for the
traceback method, i.e., the half mass radius of the ONC (1.2\,pc) compared to the
2.5\,pc boundary used in their work.
Two of their walkaway and two of their runaway stars are among our top ranked
candidates. These walkaways are sources aI+1, which is our fastest source that
passes all the flags (their estimated age for this source is $5^{+15}_{-4}$\,Myr),
and aIII+4 that we could not evaluate on variability or YSO probability (their
estimated age for this source is $0.8^{+9}_{-0.7}$\,Myr). The coinciding runaway
candidates are sources aI+*1 and aII+*1, that fulfill all quality criteria, but
that we note are in a zone contaminated by Galactic streaming.  Based on age, they
have flagged as candidates two sources that we confirm likely come from the ONC
based on 3D traceback, these are sources aIII+1 and aIII+*4, for which they have
estimated ages larger than 20\,Myr. 

The majority of the sources presented here, together with most of the candidates
presented by \cite{Schoettler2020}, have missing radial velocities. One of the most
important follow-up observations stimulated by this work thus involves radial
velocity measurements of our candidate runaways to confirm their ejection with the
more restrictive 3D traceback criterion. 

We remark that one important goal of finding the oldest ejected runaways, which may
be currently hidden among a cloud of Galactic field contaminants, is to constrain
the star formation history of clusters, i.e., the timescale of star cluster
formation. For the ONC, this runaway age constraint is still set by \mucol\ and
\aeaur, ejected about 2.5\,Myr ago \citep{Hoogerwerf2001}. The formation time is
obviously a basic parameter for cluster formation theories
\citep[e.g.][]{Tan2006,Nakamura2007}  and also influences estimates of fundamental
star formation properties, such as the efficiency per freefall time \citep[e.g.,
][]{DaRio2014}.  The cluster formation time is also directly related to the
formation timescale of massive star formation in competitive accretion models
\citep{Wang2010} as discussed by \cite{Tan2014}. Thus a dedicated search to find
relatively old, likely lower-mass runaways from the ONC should be attempted, with a
starting point from our main 2D traceback sample.

\acknowledgements 
JPF and JCT acknowledge support from ERC Advanced Grant MSTAR.
This work has made use of data from the European Space Agency (ESA) mission
\gaia\ (\url{https://www.cosmos.esa.int/gaia}), processed by the \gaia\
Data Processing and Analysis Consortium (DPAC,
\url{https://www.cosmos.esa.int/web/gaia/dpac/consortium}). Funding for the DPAC
has been provided by national institutions, in particular the institutions
participating in the \gaia\ Multilateral Agreement.

\bibliography{library.bib}

\end{document}